\documentclass[a4paper]{article}
\usepackage{amsfonts,amssymb,amsmath,amsthm,cite}
\usepackage{ucs} 
\usepackage[utf8x]{inputenc}
\usepackage{graphicx}
\usepackage[english]{babel}
\usepackage{slashed}
\usepackage{textcomp}
\usepackage{bm}
\usepackage{titlesec}
\usepackage{stackengine}

\usepackage{etoolbox}
\patchcmd{\thebibliography}{\section*}{\section}{}{}
\usepackage{adjustbox}
\usepackage{nccmath}
\usepackage{mathtools}
\mathtoolsset{showonlyrefs}
\evensidemargin=\oddsidemargin
\begin{document}
%\begin{flushright}
%	%\textbf{PDMI Preprint -- 02/2025}
%\end{flushright}
\vspace{10mm}
\begin{center}
	\large{\textbf{Four-loop renormalization with a cutoff }\\ 
		\textbf{in a sextic model}}
\end{center}
\vspace{2mm}
\begin{center}
	\large{\textbf{N. V. Kharuk}}
\end{center}
\begin{center}
	St. Petersburg Department of Steklov Mathematical Institute of Russian Academy of Sciences, 27 Fontanka, St. Petersburg, 191023, Russia
\end{center}

\begin{center}
	E-mail: natakharuk@mail.ru
\end{center}

\vspace{10mm}

\textbf{Abstract.} The quantum action for a three-dimensional real sextic model using the background field method is considered. Four-loop renormalization of this model is performed with a cutoff regularization in the coordinate representation. The coefficients for the renormalization constants are found, the applicability of the $\mathcal{R}$-operation within the proposed regularization is explicitly demonstrated, and the absence of nonlocal contributions is proved. Additionally, the explicit form of the singularities, power and logarithmic, as well as their dependence on the deformation of the Green's function are discussed.

\vspace{2mm}

\tableofcontents

\newpage

\section{Introduction}

The perturbative approach is a widely used and powerful tool in modern quantum field theory. However, its use involves the challenge of dealing with divergent integrals. To address this issue, regularization techniques are introduced into the theory. There are several popular regularization schemes used in practice, such as dimensional regularization \cite{19,555}, higher derivative regularization \cite{Bakeyev-Slavnov,29-st}, and Pauli--Villars regularization \cite{Pauli-Villars}.

This article discusses a cutoff regularization, which is a natural approach to dealing with divergent integrals. In the standard formulation, this involves cutting off  limits (or domain) of integration \cite{3,105}. We use this scheme in the form proposed in \cite{34,Ivanov-Kharuk-2020,Ivanov-Kharuk-20222,Ivanov-Akac,Ivanov-Kharuk-2023,Iv-2024-1,Iv-Kh-2024,Kh-2024,Ivanov-2022,Iv-2024} to find four-loop divergences in the sextic scalar model \cite{NK-1,NK-2,NK-3,NK-4,NK-5,NK-6}. A distinguishing feature of our approach is not simply cutting the integration limits but rather a specific deformation of the Green's function, which allows us to maintain quasi-locality in the theory. This advantage is significant because it makes the regularization method more widely applicable, particularly when working with smooth compact manifolds and using gluing techniques \cite{sk-14,sk-16,sk-7,sksk}.

In this paper, we use the background field method to analyze divergences \cite{102,103,24,25,26}. This method is convenient as it allows us to check all necessary diagrammatic relations while working with only one object, an effective action. Our choice of sextic model is justified both by its relative simplicity from the mathematical and computational standpoints, and by the significant physical role it plays \cite{nknk-1,nknk-2}. The main result can be divided into four parts.
\begin{itemize}
    \item The fourth coefficients of the renormalization constants have been calculated.
	\item The applicability of the $\mathcal{R}$-operation has been verified. 
	\item Nonlocal contributions have been shown to be absent.
	\item The dependence of singular contributions on the type of deformation of the Green's function has been studied.
\end{itemize}

Note that this is the first instance of four-loop renormalization using the cutoff regularization in the coordinate representation. This study can be viewed as a continuation of a series of papers on scalar theories with the cutoff regularization \cite{34,Iv-2024-1,Iv-Kh-2024}, and in particular, as an extension of the study on the sextic model \cite{Kh-2024}.

The paper has the following structure. Section \ref{NK-s-p} introduces the problem statement, basic objects, and definitions. Section \ref{NK-s-r} presents the main results. Section \ref{NK-s-v} provides detailed calculations and auxiliary relations. Section \ref{NK:sec:dis} is the final section, containing brief conclusions and further discussion.

\section{The problem statement}
\label{NK-s-p}

Consider the 3-dimensional Euclidean space $\mathbb{R}^3$. The elements of this space are notated by Latin letters  $x, y, z$, and their individual components are indexed using Greek letters. Additionally, we introduce a scalar real field $\phi(\cdot)$ and a classical action $S[\,\cdot\,]$ for a model with sextic interaction
\begin{equation}\label{NK-f-1}
S_{\mathrm{cl}}[\phi]=S_0[\phi]+S_{\mathrm{int}}[\phi],
\end{equation}
where
\begin{equation}\label{NK-f-2}
S_0[\phi]=\frac{1}{2}\int_{\mathbb{R}^3}\mathrm{d}^3x\,
\phi(x)A_0(x)\phi(x),\,\,\,
S_{\mathrm{int}}[\phi]=\int_{\mathbb{R}^3}\mathrm{d}^3x\bigg(
\frac{m^2}{2}\phi^2(x)+\frac{t_4}{4!}\phi^4(x)+\frac{t_6}{6!}\phi^6(x)\bigg).
\end{equation}
Here, $A_0(x)=-\partial_{x_\mu}\partial_{x^\mu}$ is the standard Laplacian operator, $m^2$ is the squared mass parameter, and $t_4$ and $t_6$ are interaction constants. In the context of the perturbative approach, their real values are not relevant. We assume $\Re(t_6) > 0$ to make functional $S_{\mathrm{int}}[\,\cdot\,]$ be bounded from below. Note that this type of model only includes even powers. The variant that includes odd powers was studied in \cite{Kh-2024} as part of three-loop renormalization, considering the arbitrariness of finite terms as well.

As is well known, the quantum action for such a model includes ultraviolet divergences. Therefore, it is necessary to introduce a regularization and then perform renormalization \cite{6,7}. In this study, regularization  refer to the following type of deformation
\begin{equation}\label{NK-f-3}
S_{\mathrm{int}}[\phi]\to S_{\mathrm{int}}[\phi^\Lambda_\omega],
\end{equation}
where
\begin{equation}\label{NK-f-4}
\phi^\Lambda_\omega=\int_{\mathbb{R}^3}\mathrm{d}^3y\,\omega(|y|)\phi(x+y/\Lambda),\,\,\,
\int_{\mathbb{R}^3}\mathrm{d}^3y\,\omega(|y|)=1,\,\,\,\mathrm{supp}(\omega)\subset[0,1].
\end{equation}
In the last line $\Lambda$ is a regularizing parameter. In the limit $\Lambda\to+\infty$ the regularization is removed. This type of regularization leads to the replacement of the Green's function for the free Laplace operator
\begin{equation}\label{NK-f-5}
R_0^{\phantom{1}}(x)=\frac{1}{4\pi|x|}\to R_0^\Lambda(x)=\frac{1}{4\pi}
\begin{cases}
\Lambda\big(1+\mathbf{f}(|x|^2\Lambda^2)\big), & |x|\leqslant1/\Lambda;\\
\,\,\,\,\,\,\,\,\,\,\,\,\,\,\,\,|x|^{-1},& |x|>1/\Lambda,
\end{cases}
\end{equation} 
in all elements of the perturbative expansion. In the last formula, the function $\mathbf{f}$ belongs to $C([0,1],\mathbb{R})$ and has the property $\mathbf{f}(1)=0$. Note that the formulation of the regularization \eqref{NK-f-3} is equivalent to a special deformation of the operator $A_0^{\phantom{1}}(x)\to A_0^\Lambda(x)$, which was used in \cite{Kh-2024,Iv-2024-1,Iv-Kh-2024} within the framework of three-loop calculations.

In the case of the sextic model in three-dimensional space, multiplicative renormalization is applicable, which consists in redefining the parameters of the theory
\begin{equation}\label{NK-f-6}
\phi(\cdot)\to\phi(\cdot)Z_0^{1/2},\,\,\,
m^2\to m^2Z_2/Z_0,\,\,\,
t_4\to t_4Z_4/Z_0^{2},\,\,\,t_6\to t_6Z_6/Z_0^{3},
\end{equation}
where
\begin{equation}\label{NK-f-7}
Z_n=z_{n0}+\sum_{k=1}^{+\infty}\hbar^kz_{nk}\,\,\,\mbox{and}\,\,\,z_{n0}=1\,\,\,\mbox{for all}\,\,\,n\in\{0,2,4,6\}.
\end{equation}

After applying the background field method $\phi(\cdot)\to \tilde{B}(\cdot)+\sqrt{\hbar}\phi(\cdot)$, described in detail in \cite{102,103,24,25,26,Iv-2024-1}, the perturbative expansion for the quantum action can be written explicitly. In what follows, the background field $B(\cdot)$ denotes the deformed $\tilde{B}_\omega^\Lambda(\cdot)$. For the formulation, we introduce several additional auxiliary functionals
\begin{equation}\label{NK-f-8}
\mathrm{V}_{i,j}\equiv\mathrm{V}_{i,j}[\phi,B]=\int_{\mathbb{R}^3}\mathrm{d}^3x\,
\big(\phi_\omega^\Lambda(x)\big)^iB^j(x),\,\,\,\mbox{where}\,\,\,
i,j\in\mathbb{N}\cup\{0\}\,\,\,\mbox{and}\,\,\,i+j>0,
\end{equation}
\begin{equation}\label{NK-f-10}
\Gamma_{3k}[\phi]=t_4z_{4k}\mathrm{V}_{3,1}+
\frac{t_6z_{6k}}{3!}\mathrm{V}_{3,3},\,\,\,
\Gamma_{4k}[\phi]=t_4z_{4k}\mathrm{V}_{4,0}+
\frac{t_6z_{6k}}{2}\mathrm{V}_{4,2},
\end{equation}
\begin{equation}\label{NK-f-12}
	\Gamma_{5k}[\phi]=t_6z_{6k}\mathrm{V}_{5,1},\,\,\,
	\Gamma_{6k}[\phi]=t_6z_{6k}\mathrm{V}_{6,0},
\end{equation}
\begin{equation}\label{NK-f-11}
\mathrm{X}_i[\phi]=2z_{0i}S_0[\phi]+m^2z_{2i}\mathrm{V}_{2,0}+\frac{t_4z_{4i}}{2}\mathrm{V}_{2,2}+\frac{t_6z_{6i}}{4!}\mathrm{V}_{2,4}.
\end{equation}
We also notate by the symbol $G^\Lambda(x,y)$ the Green's function for the quadratic form  operator $\mathrm{X}_0[\phi]$. As is known, the expansion for such a function near the diagonal, see \cite{29,30-1-1}, is written out explicitly
\begin{equation}\label{NK-f-z10}
G^\Lambda(x,y)=R_0^\Lambda(x-y)-g_\Lambda(x-y)\bigg(m^2+t_4\frac{B^2(x)+B^2(y)}{4}+t_6\frac{B^4(x)+B^4(y)}{48}\bigg)+PS(x,y),
\end{equation}
where
\begin{equation}\label{NK-f-z12}
g_\Lambda(x-y)=
\int_{\mathrm{B}_{1/\sigma}}\mathrm{d}^3z\,R_0^\Lambda(x-y+z)R_0^\Lambda(z)-
\int_{\mathrm{B}_{1/\sigma}}\mathrm{d}^3z\,\Big(R_0^\Lambda(z)\Big)^2,
\end{equation}
and $PS(x,y)$ is a nonlocal component, which has two finite derivatives.
In particular, we have
\begin{equation}\label{NK-f-z11}
G^\Lambda(x,x)=R_0^\Lambda(0)+PS(x,x).
\end{equation}
Note that for functionals and the Green's function, elements of diagram technique can be introduced in the form
\begin{equation}\label{NK-dd-1}
G^\Lambda={\centering\adjincludegraphics[width = 1.2 cm, valign=c]{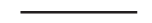}},\,\,\,
\Gamma_{30}={\centering\adjincludegraphics[width = 1.2 cm, valign=c]{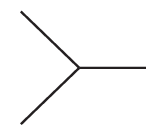}},\,\,\,
\Gamma_{40}={\centering\adjincludegraphics[width = 1.2 cm, valign=c]{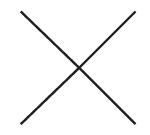}},\,\,\,
\Gamma_{50}={\centering\adjincludegraphics[width = 1.2 cm, valign=c]{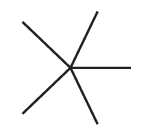}},\,\,\,
\Gamma_{60}={\centering\adjincludegraphics[width = 1.2 cm, valign=c]{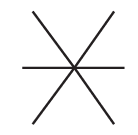}},
\end{equation}
\begin{equation}\label{NK-dd-2}
\mathrm{X}_{i}={\centering\adjincludegraphics[width = 1.7 cm, valign=c]{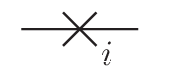}},\,\,\,
\Gamma_{31}={\centering\adjincludegraphics[width = 1.2 cm, valign=c]{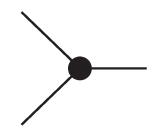}},\,\,\,
\Gamma_{41}={\centering\adjincludegraphics[width = 1.2 cm, valign=c]{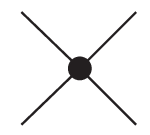}},
\end{equation}
\begin{equation}\label{NK-dd-3}
\Gamma_{32}={\centering\adjincludegraphics[width = 1.2 cm, valign=c]{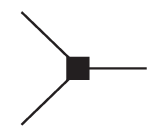}},\,\,\,
\Gamma_{42}={\centering\adjincludegraphics[width = 1.2 cm, valign=c]{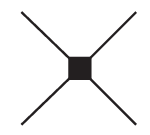}}.
\end{equation}
Then the quantum renormalized action $W_{\mathrm{ren}}[B,\Lambda]$ can be written as follows
\begin{multline}\label{NK-f-9}
	W_{\mathrm{ren}}[B,\Lambda]=\int_{\mathbb{R}^3}\mathrm{d}^3x\bigg(
	\frac{z_{04}}{2}\tilde{B}(x)A_0(x)\tilde{B}(x)+\frac{z_{24}m^2}{2}B^2(x)+\frac{z_{44}t_4}{4!}B^4(x)+\frac{z_{64}t_6}{6!}B^6(x)\bigg)-\\
	-\bigg(\frac{\hbar}{2}\ln\det(G^\Lambda)+\hbar\kappa_1\bigg)
	-\Bigg[\hbar\exp\Bigg(-\frac{1}{2}\sum_{k=1}^{+\infty}\hbar^k \mathrm{X}_k[\delta_j]-\sum_{n=3}^6\sum_{k=0}^{+\infty}\frac{\hbar^{n/2+k-1}}{n!}\Gamma_{nk}[\delta_j]\Bigg)\times\\\times e^{g[G^\Lambda,j]}\bigg|^{1\mathrm{PI}}_{j=0}+\sum_{n=2}^{+\infty}\hbar^n\kappa_n\Bigg],
\end{multline}
where $j(x)$ is an auxiliary smooth field, $\delta_{j(x)}$ is the variational derivative with respect to the field $j(x)$, and
\begin{equation*}\label{NK-f-13}
	g[G^\Lambda,j]=\frac{1}{2}\int\int_{\mathbb{R}^3\times\mathbb{R}^3}\mathrm{d}^3x\mathrm{d}^3y\,j(x)G^\Lambda(x,y)j(y).
\end{equation*}
Also, the symbol <<$1\mathrm{PI}$>> means that only strongly connected diagrams are preserved in the sum, they are also called one-particle irreducible. The constants $\kappa_n$ subtract singularities that do not depend on the background field.

Given that the subtraction of singularities can be performed by a suitable choice of the coefficients of the renormalization constants \eqref{NK-f-7}, the perturbative expansion \eqref{NK-f-9} defines a set of equations that uniquely determine the singular components of the desired coefficients. Earlier in \cite{Kh-2024}, the first three relations were investigated, which allowed renormalization in three loops. The coefficients found in the minimal subtraction scheme, which is an analogue of the MS-scheme for the dimensional regularization, have the form
\begin{equation*}
	Z_0=1+\mathcal{O}\big(\hbar^4\big),
\end{equation*}
\begin{equation*}
	Z_2=1-\hbar\Lambda\frac{\alpha t_4}{2m^2}+
	\hbar^2\bigg(
	\Lambda^2\frac{\alpha^2t_6}{8m^2}
	+L\frac{t_4^2}{96\pi^2m^2}
	\bigg)+\hbar^3\bigg(
	-\Lambda L
	\frac{\alpha t_4t_6}{48\pi^2m^2}
	+
	\Lambda\frac{\alpha_1(\mathbf{f})t_4t_6}{24m^2}
	\bigg)+\mathcal{O}\big(\hbar^4\big),
\end{equation*}
\begin{equation*}
	Z_4=1-\hbar\Lambda\frac{\alpha t_6}{2t_4}+
	\hbar^2L\frac{t_6}{24\pi^2}+\hbar^3
	\bigg(
	-\Lambda L
	\frac{5\alpha t_6^2}{96\pi^2t_4}
	+
	\Lambda\frac{\alpha_1(\mathbf{f})t_6^2}{8t_4}
	\bigg)+\mathcal{O}\big(\hbar^4\big),
\end{equation*}
\begin{equation*}
	Z_6=1+\hbar^2L\frac{5t_6}{48\pi^2}+\mathcal{O}\big(\hbar^4\big).
\end{equation*}
The aim of this work is to study the fourth (proportional to $\hbar^4$) relation and to find the coefficients $z_{04},z_{24},z_{44},z_{64}$ for the renormalization constants.

\section{Results}
\label{NK-s-r}

%\begin{theorem}\label{nk-t-1}
Taking into account all the above, the part of the renormalized effective action \eqref{NK-f-9} proportional to $\hbar^4$ does not contain nonlocal singular contributions, i.e., those depending on the function $PS(x,y)$. The remaining singularities are cancelled by the following choice of coefficients for the renormalization constants
\begin{equation}\label{NK-f1-1}
z_{04}=-L\frac{t^2_6}{2^43^25(4\pi)^4},
\end{equation} 
\begin{equation}\label{1}
z_{24}=
L\Lambda^2\frac{5\alpha^2(\mathbf{f})t_6^2}{2^33(4\pi)^2m^2}
-\Lambda^2\frac{t^2_6(15\alpha_1(\mathbf{f})\alpha(\mathbf{f})-2\alpha_2(\mathbf{f}))}{5!2m^2}+
L^2\frac{t_4^2t_6^{\phantom{1}}}{3^2(4\pi)^4m^2}-L\frac{t_4^2t_6^{\phantom{1}}(32+3\pi^2)-4m^2t^2_6}{2^63(4\pi)^4m^2},
\end{equation} 
\begin{equation}\label{NK-f-z3}
z_{44}=L^2\frac{7t_6^2}{3^2(4\pi)^4}-L\frac{t_4t_6^2(116+9\pi^2)}{2^53(4\pi)^4},
\end{equation}
\begin{equation}\label{NK-f-z4}
z_{64}=L^2\frac{25t_6^2}{3^2(4\pi)^4}-L\frac{t_6^2(150+15\pi^2)}{2^5(4\pi)^4},
\end{equation}
where auxiliary numbers have the form
\begin{equation}\label{NK-f-z6}
\alpha(\mathbf{f})=\frac{\mathbf{f}(0)+1}{4\pi},
\end{equation}
\begin{equation}\label{NK-f-z7}
\alpha_1(\mathbf{f})=\int_{\mathbb{R}^3}\mathrm{d}^3y\,\Big(R_0^1(y)\Big)^4=
\frac{1}{(4\pi)^3}\bigg(1+\int_0^1\mathrm{d}t\,t^2\Big(\mathbf{f}(t^2)+1\Big)^4\bigg),
\end{equation}
\begin{equation}\label{NK-f-z8}
\alpha_2(\mathbf{f})=\int_{\mathbb{R}^3}\mathrm{d}^3y\,\Big(R_0^1(y)\Big)^5=
\frac{1}{(4\pi)^4}\bigg(\frac{1}{2}+\int_0^1\mathrm{d}t\,t^2\Big(\mathbf{f}(t^2)+1\Big)^5\bigg).
\end{equation}
%\begin{equation}\label{NK-f-z5}
%\alpha_3=
%\int_{\mathbb{R}^3}\mathrm{d}^3x
%\int_{|\hat{y}|=1}\mathrm{d}^2\sigma(\hat{y})\,\Big(R_0(x)\Big)^2
%\Big(R_0(x-\hat{y})\Big)^2.
%\end{equation}
%\end{theorem}
%\begin{proof}
 All the basic calculations are given in section \ref{NK-s-v}, so here we  only note the main stages of the process. First, it is necessary to write out the four-loop relation in order to find the coefficients. It has the form \eqref{NK-f-z9}. Next, it is necessary to find the singular components on the right-hand side of the equality. It is more convenient to analyze them by groups. For this purpose, the right-hand side is divided into 17 parts, for each of which separate calculations are carried out, see section \ref{kn-sec-sot}. Then the answers are summed up. As a result, only the parts of the classical action remain, see formulas \eqref{NK-f-62}, \eqref{NK-f-65}, \eqref{NK-f-68-1}, and \eqref{NK-f-73}, multiplied by the singular coefficients. Finally, the coefficients $z_{04},z_{24},z_{44},z_{64}$ on the left-hand side of equality \eqref{NK-f-z9} are selected in such a way that the singular components on both sides of the equality coincide. This leads to the obtained coefficients.

Separately, we note that the calculated coefficients are consistent with the results obtained using the dimensional regularization \cite{NK-5}.
%\end{proof}

\section{Calculations}
\label{NK-s-v}

\subsection{Auxiliary expansions}
\textbf{Definition:} Let $n,j,i\in\mathbb{N}\cup\{0\}$, $j\leqslant n$, $\Omega[\phi]$ be a functional proportional to the $n$-th power of the field, that is, $\Omega[s\phi]=s^n\Omega[\phi]$ for $s>0$. Let us introduce several operators into consideration.
\begin{itemize}
	\item The operator $\mathbb{H}_{j,i}^{\mathrm{c\,(sc)}}$ transforms the functional $\Omega[\phi]$ into a functional proportional to the $j$-th power of the field, by means of all possible pairings of $n-j$ functions of the field $\phi(\cdot)$ using the regularized Green's function $G^\Lambda(\,\cdot\,,\,\cdot\,)$, that is, using substitutions of the form $\phi(x)\phi(y)\to G^\Lambda(x,y)$, and preserving only the connected (strongly connected) part containing $i$ Green's functions on the diagonal (loops).
	\item The operator without projection on the number of loops has the form $$\mathbb{H}_{j}^{\mathrm{c\,(sc)}}=\sum_{i=0}^{+\infty}\mathbb{H}_{j,i}^{\mathrm{c\,(sc)}}.$$
\end{itemize}
Given the last definition, see the analog in \cite{I-R}, the relation for finding the fourth coefficients, that is, the part in  formula \eqref{NK-f-9} proportional to $\hbar^4$, can be represented as follows
\begin{multline}\label{NK-f-z9}
\int_{\mathbb{R}^3}\mathrm{d}^3x\bigg(
\frac{Z_0}{2}\tilde{B}(x)A_0(x)\tilde{B}(x)+\frac{Z_2m^2}{2}B^2(x)+\frac{Z_4t_4}{4!}B^4(x)+\frac{Z_6t_6}{6!}B^6(x)\bigg)\stackrel{\mathrm{s.p.}}{=}\\
\stackrel{\mathrm{s.p.}}{=}\frac{1}{4!}\frac{\mathrm{d}^4}{\mathrm{d}\hbar^4}\Bigg|_{\hbar=0}\mathbb{H}_0^{\mathrm{sc}}\Bigg(\hbar\exp\Bigg(-\frac{1}{2}\sum_{k=1}^{+\infty}\hbar^k \mathrm{X}_k[\phi]-\sum_{n=3}^6\sum_{k=0}^{+\infty}\frac{\hbar^{n/2+k-1}}{n!}\Gamma_{nk}[\phi]\Bigg)\Bigg)+\kappa_4,
\end{multline}
where the sign $\stackrel{\mathrm{s.p.}}{=}$ means the equality of singular components (by parameter $\Lambda$). The left side of the last equality can be rewritten as several contributions. We write them out separately with additional comments. The first component is the usual (without counter-terms) diagrams
\begin{multline}\label{NK-f-14}
\frac{\mathbb{H}_0^{\mathrm{sc}}\big(\Gamma_{30}^6\big)}{(3!)^66!}-
\frac{\mathbb{H}_0^{\mathrm{sc}}\big(\Gamma_{30}^4\Gamma_{40}^{\phantom{1}}\big)}{(3!)^4(4!)^2}+
\frac{\mathbb{H}_0^{\mathrm{sc}}\big(\Gamma_{30}^3\Gamma_{50}^{\phantom{1}}\big)}{(3!)^45!}+
\frac{\mathbb{H}_0^{\mathrm{sc}}\big(\Gamma_{30}^2\Gamma_{40}^2\big)}{4(3!)^2(4!)^2}-\\-
\frac{\mathbb{H}_0^{\mathrm{sc}}\big(\Gamma_{30}^2\Gamma_{60}^{\phantom{1}}\big)}{2(3!)^26!}-
\frac{\mathbb{H}_0^{\mathrm{sc}}\big(\Gamma_{30}^{\phantom{1}}\Gamma_{40}^{\phantom{1}}\Gamma_{50}^{\phantom{1}}\big)}{3!4!5!}
-\frac{\mathbb{H}_0^{\mathrm{sc}}\big(\Gamma_{40}^3\big)}{3!(4!)^3}+
\frac{\mathbb{H}_0^{\mathrm{sc}}\big(\Gamma_{50}^2\big)}{2(5!)^2}+\frac{\mathbb{H}_0^{\mathrm{sc}}\big(\Gamma_{40}^{\phantom{1}}\Gamma_{60}^{\phantom{1}}\big)}{4!6!}.
\end{multline}
The second component is the counter-term diagrams, which are obtained from the three-loop diagrams either by adding $\mathrm{X}_1$ or by replacing one of the vertices $\Gamma_{30}$--$\Gamma_{60}$ with $\Gamma_{31}$--$\Gamma_{61}$. They have the following form
\begin{multline}\label{NK-f-15}
	\frac{\mathbb{H}_0^{\mathrm{sc}}\big(\Gamma_{30}^3\Gamma_{31}^{\phantom{1}}\big)}{(3!)^5}-
	\frac{\mathbb{H}_0^{\mathrm{sc}}\big(\Gamma_{30}^{\phantom{1}}\Gamma_{31}^{\phantom{1}}\Gamma_{40}^{\phantom{1}}\big)}{(3!)^24!}-
	\frac{\mathbb{H}_0^{\mathrm{sc}}\big(\Gamma_{30}^2\Gamma_{41}^{\phantom{1}}\big)}{2(3!)^24!}+
	\frac{\mathbb{H}_0^{\mathrm{sc}}\big(\Gamma_{40}^{\phantom{1}}\Gamma_{41}^{\phantom{1}}\big)}{(4!)^2}+
	\frac{\mathbb{H}_0^{\mathrm{sc}}\big(\Gamma_{31}^{\phantom{1}}\Gamma_{50}^{\phantom{1}}\big)}{3!5!}-\\-
	\frac{\mathbb{H}_0^{\mathrm{sc}}\big(\Gamma_{30}^4\mathrm{X}_1^{\phantom{1}}\big)}{2(3!)^44!}+
	\frac{\mathbb{H}_0^{\mathrm{sc}}\big(\Gamma_{30}^2\Gamma_{40}^{\phantom{1}}\mathrm{X}_1^{\phantom{1}}\big)}{4(3!)^24!}-
	\frac{\mathbb{H}_0^{\mathrm{sc}}\big(\Gamma_{40}^2\mathrm{X}_1^{\phantom{1}}\big)}{4(4!)^2}-
	\frac{\mathbb{H}_0^{\mathrm{sc}}\big(\Gamma_{30}^{\phantom{1}}\Gamma_{50}^{\phantom{1}}\mathrm{X}_1^{\phantom{1}}\big)}{2(3!5!)}+
	\frac{\mathbb{H}_0^{\mathrm{sc}}\big(\Gamma_{60}^{\phantom{1}}\mathrm{X}_1^{\phantom{1}}\big)}{2(6!)}.
\end{multline}
The third component is the counter-term diagrams, which are obtained from the two-loop diagrams either by adding $\mathrm{X}_2$ or $\mathrm{X}_1\mathrm{X}_1$, or by replacing one of the vertices $\Gamma_{30}$--$\Gamma_{60}$ with $\Gamma_{32}$--$\Gamma_{62}$, or by replacing two vertices $\Gamma_{30}$--$\Gamma_{60}$ with $\Gamma_{31}$--$\Gamma_{61}$, or by replacing one vertex $\Gamma_{30}$--$\Gamma_{60}$ with $\Gamma_{31}$--$\Gamma_{61}$ and adding $\mathrm{X}_1$. They have the following form
\begin{multline}\label{NK-f-16}
-\frac{\mathbb{H}_0^{\mathrm{sc}}\big(\Gamma_{30}^2\mathrm{X}_2^{\phantom{1}}\big)}{4(3!)^2}+
\frac{\mathbb{H}_0^{\mathrm{sc}}\big(\Gamma_{40}^{\phantom{1}}\mathrm{X}_2^{\phantom{1}}\big)}{2(4!)}
+\frac{\mathbb{H}_0^{\mathrm{sc}}\big(\Gamma_{30}^2\mathrm{X}_1^2\big)}{2^4(3!)^2}-
\frac{\mathbb{H}_0^{\mathrm{sc}}\big(\Gamma_{40}^{\phantom{1}}\mathrm{X}_1^2\big)}{2^3(4!)}+\\+
\frac{\mathbb{H}_0^{\mathrm{sc}}\big(\Gamma_{30}^{\phantom{1}}\Gamma_{32}^{\phantom{1}}\big)}{(3!)^2}-
\frac{\mathbb{H}_0^{\mathrm{sc}}\big(\Gamma_{42}^{\phantom{1}}\big)}{4!}+\frac{\mathbb{H}_0^{\mathrm{sc}}\big(\Gamma_{31}^2\big)}{2(3!)^2}
-\frac{\mathbb{H}_0^{\mathrm{sc}}\big(\Gamma_{30}^{\phantom{1}}\Gamma_{31}^{\phantom{1}}\mathrm{X}_1^{\phantom{1}}\big)}{2(3!)^2}+
\frac{\mathbb{H}_0^{\mathrm{sc}}\big(\Gamma_{41}^{\phantom{1}}\mathrm{X}_1^{\phantom{1}}\big)}{2(4!)}.
\end{multline}
Finally, the fourth component is the combination of vertices $\mathrm{X}_1$--$\mathrm{X}_3$ of the form
\begin{equation}\label{NK-f-17}
-\frac{\mathbb{H}_0^{\mathrm{sc}}\big(\mathrm{X}_1^3\big)}{2^3(3!)}+
\frac{\mathbb{H}_0^{\mathrm{sc}}\big(\mathrm{X}_1^{\phantom{1}}\mathrm{X}_2^{\phantom{1}}\big)}{4}-
\frac{\mathbb{H}_0^{\mathrm{sc}}\big(\mathrm{X}_3^{\phantom{1}}\big)}{2}.
\end{equation}

\subsection{Types of divergent diagrams}
The search for singular components is based on the $\mathcal{R}$-operation \cite{Bog-R}, which allows one to remove internal divergences from a diagram. In this paper, an explicit calculation is given, showing that within the framework of the proposed regularization, the basic rules for working with subdivergences remain true. Note that for an arbitrary type of regularization, this well-known approach is not transparent and the feasibility of its application requires additional research.

Below we present all diagrams from \eqref{NK-f-14} that contain singularities. Some terms that are finite are notated by ellipsis. The term $\mathbb{H}_0\big(\Gamma_{30}^6\big)$ is completely absent, because in this case strongly connected diagrams do not contain the Green's functions on the diagonal and more than two identical lines. The remaining equalities are conveniently presented as follows.
\begin{fleqn}
\begin{gather}\label{NK-ff-1}
\mathbb{H}_0^{\mathrm{sc}}\big(\Gamma_{40}^3\big)=
3^32^6\mathrm{A}_{27}+
3^42^5\mathrm{A}_{26}+
3^32^7\mathrm{A}_{25}+
3^32^6\mathrm{A}_{24},
\end{gather}
\end{fleqn}
\begin{equation*}
\mathrm{A}_{27}={\centering\adjincludegraphics[width = 1.5 cm, valign=c]{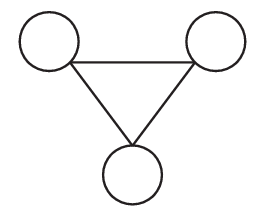}},\,\,\,
\mathrm{A}_{26}={\centering\adjincludegraphics[width = 2 cm, valign=c]{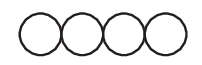}},\,\,\,
\mathrm{A}_{25}={\centering\adjincludegraphics[width = 1.7 cm, valign=c]{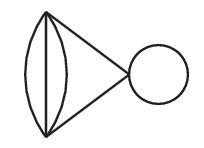}},\,\,\,
\mathrm{A}_{24}={\centering\adjincludegraphics[width = 1.6 cm, valign=c]{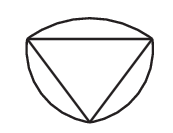}},\hfill
\end{equation*}
\begin{fleqn}
\begin{gather}\label{NK-ff-2}
\mathbb{H}_0^{\mathrm{sc}}\big(\Gamma_{60}\Gamma_{40}\big)=
5^13^32^2\mathrm{A}_{23}+
5^13^22^3\mathrm{A}_{22},
\end{gather}
\end{fleqn}
\begin{equation*}
\mathrm{A}_{23}={\centering\adjincludegraphics[width = 1.9 cm, valign=c]{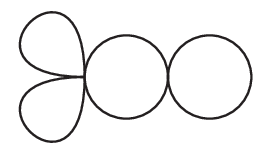}},\,\,\,
\mathrm{A}_{22}={\centering\adjincludegraphics[width = 1.8 cm, valign=c]{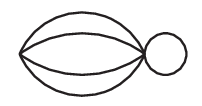}},
\end{equation*}
\begin{fleqn}
\begin{gather}\label{NK-ff-3}
\mathbb{H}_0^{\mathrm{sc}}\big(\Gamma_{50}^2\big)=
5^13^12^3\mathrm{A}_{21}+
5^23^12^3\mathrm{A}_{20},
\end{gather}
\end{fleqn}
\begin{equation*}
\mathrm{A}_{21}={\centering\adjincludegraphics[width = 1.5 cm, valign=c]{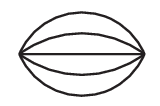}},\,\,\,
\mathrm{A}_{20}={\centering\adjincludegraphics[width = 2.3 cm, valign=c]{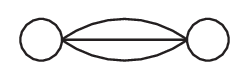}},
\end{equation*}
\begin{fleqn}
\begin{gather}\label{NK-ff-4}
\mathbb{H}_0^{\mathrm{sc}}\big(\Gamma_{60}^{\phantom{1}}\Gamma_{30}^2\big)=
5^13^22^4\mathrm{A}_{19}+
5^13^42^3\mathrm{A}_{18}+
5^13^42^2\mathrm{A}_{17},
\end{gather}
\end{fleqn}
\begin{equation*}
\mathrm{A}_{19}={\centering\adjincludegraphics[width = 1.8 cm, valign=c]{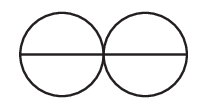}},\,\,\,
\mathrm{A}_{18}={\centering\adjincludegraphics[width = 1.4 cm, valign=c]{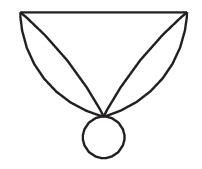}},\,\,\,
\mathrm{A}_{17}={\centering\adjincludegraphics[width = 1.8 cm, valign=c]{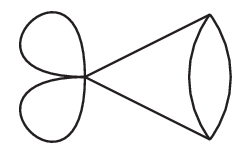}},
\end{equation*}
\begin{fleqn}
\begin{gather}\label{NK-ff-5}
\mathbb{H}_0^{\mathrm{sc}}\big(\Gamma_{50}^{\phantom{1}}\Gamma_{30}^3\big)=
5^13^52^5\mathrm{A}_{16}+
5^13^42^5\mathrm{A}_{15}+
5^13^42^4\mathrm{A}_{14}+\ldots,
\end{gather}
\end{fleqn}
\begin{equation*}
\mathrm{A}_{16}={\centering\adjincludegraphics[width = 1.4 cm, valign=c]{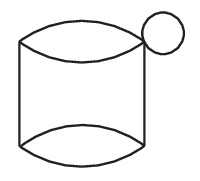}},\,\,\,
\mathrm{A}_{15}={\centering\adjincludegraphics[width = 1.2 cm, valign=c]{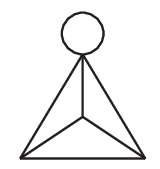}},\,\,\,
\mathrm{A}_{14}={\centering\adjincludegraphics[width = 2 cm, valign=c]{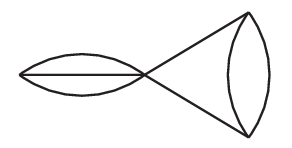}},
\end{equation*}
\begin{fleqn}
\begin{gather}\label{NK-ff-6}
\mathbb{H}_0^{\mathrm{sc}}\big(\Gamma_{50}\Gamma_{40}\Gamma_{30}\big)=
5^13^22^4\mathrm{A}_{13}+
5^13^32^5\mathrm{A}_{12}+
5^13^32^4\mathrm{A}_{11}+
5^13^22^5\mathrm{A}_{10},
\end{gather}
\end{fleqn}
\begin{equation*}
\mathrm{A}_{13}={\centering\adjincludegraphics[width = 2.3 cm, valign=c]{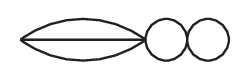}},\,\,\,
\mathrm{A}_{12}={\centering\adjincludegraphics[width = 1.8 cm, valign=c]{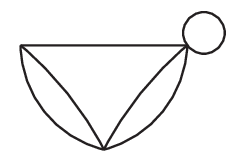}},\,\,\,
\mathrm{A}_{11}={\centering\adjincludegraphics[width = 1.4 cm, valign=c]{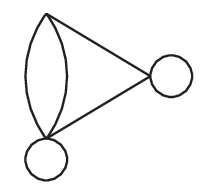}},\,\,\,
\mathrm{A}_{10}={\centering\adjincludegraphics[width = 1.4 cm, valign=c]{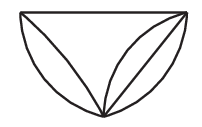}},
\end{equation*}
\begin{fleqn}
\begin{gather}\label{NK-ff-7}
\mathbb{H}_0^{\mathrm{sc}}\big(\Gamma_{40}^{\phantom{1}}\Gamma_{30}^3\big)=
3^62^6\mathrm{A}_{9}+
3^62^5\mathrm{A}_{8}+
3^62^6\mathrm{A}_{7}+\ldots,
\end{gather}
\end{fleqn}
\begin{equation*}
\mathrm{A}_{9}={\centering\adjincludegraphics[width = 1.1 cm, valign=c]{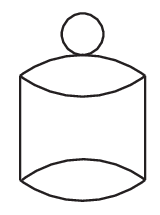}},\,\,\,
\mathrm{A}_{8}={\centering\adjincludegraphics[width = 1.4 cm, valign=c]{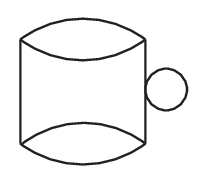}},\,\,\,
\mathrm{A}_{7}={\centering\adjincludegraphics[width = 1.4 cm, valign=c]{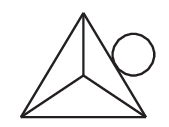}},
\end{equation*}
\begin{fleqn}
\begin{gather}\label{NK-ff-8}
\mathbb{H}_0^{\mathrm{sc}}\big(\Gamma_{40}^2\Gamma_{30}^2\big)=
3^32^7\mathrm{A}_{6}+
3^42^6\mathrm{A}_{5}+
3^42^6\mathrm{A}_{4}+
3^42^6\mathrm{A}_{3}+
3^42^7\mathrm{A}_{2}+
3^42^5\mathrm{A}_{1}+\ldots,
\end{gather}
\end{fleqn}
\begin{equation*}
\mathrm{A}_{6}={\centering\adjincludegraphics[width = 1.2 cm, valign=c]{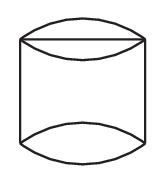}},\,\,\,
\mathrm{A}_{5}={\centering\adjincludegraphics[width = 1.9 cm, valign=c]{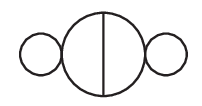}},\,\,\,
\mathrm{A}_{4}={\centering\adjincludegraphics[width = 1.5 cm, valign=c]{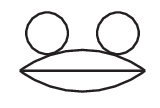}},
\end{equation*}
\begin{equation*}
	\mathrm{A}_{3}={\centering\adjincludegraphics[width = 1.8 cm, valign=c]{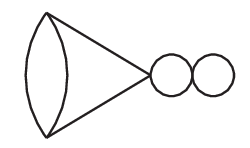}},\,\,\,
	\mathrm{A}_{2}={\centering\adjincludegraphics[width = 1.4 cm, valign=c]{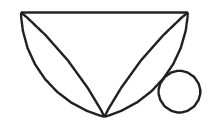}},\,\,\,
	\mathrm{A}_{1}={\centering\adjincludegraphics[width = 1.4 cm, valign=c]{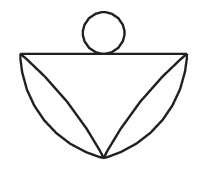}}.
\end{equation*}
\subsection{Special relations}
\label{kn-sec-sot}

This section demonstrates the relations between the diagrams presented in the previous section. In this instance, the diagrams are expressed using operator notation. This method allows for a more concise representation and greatly simplifies calculations, acting as an analogue to the well-known $\mathcal{R}$-operation within the context of the proposed regularization \eqref{NK-f-5}.\\

\noindent\textbf{Relation 1.} In the first relation, we consider all three divergent diagrams $\{\mathrm{A}_9,\mathrm{A}_8,\mathrm{A}_7\}$ from \eqref{NK-ff-7} taking into account the corresponding coefficient from \eqref{NK-f-14}. Note that all three diagrams contain a subdivergence in the form of the Green's function on the diagonal, so the selected diagrams can be uniquely expressed, separated from the other available diagrams, by replacing $\Gamma_{40}^{\phantom{1}}\to\mathbb{H}_2^{\mathrm{c}}(\Gamma_{40}^{\phantom{1}})$. Also note that the singularity in the specified subdiagram, according to the general theory, should be cancelled by the first coefficient of the renormalization constant.

This reasoning leads to a desire to consider the divergent diagrams from \eqref{NK-ff-7} together with the sixth term from \eqref{NK-f-15}, which acts as a counter-term. As a consequence,  one can ensure that an equality holds
\begin{equation}\label{NK-f-18}
-\frac{\mathbb{H}_0^{\mathrm{sc}}\big(\Gamma_{30}^4\mathbb{H}_2^{\mathrm{c}}(\Gamma_{40}^{\phantom{1}})\big)}{(3!)^4(4!)^2}
-\frac{\mathbb{H}_0^{\mathrm{sc}}\big(\Gamma_{30}^4\mathrm{X}_1^{\phantom{1}}\big)}{2(3!)^44!}=
-\frac{1}{(3!)^4(4!)^2}\mathbb{H}_0^{\mathrm{sc}}\Big(\Gamma_{30}^4\big(\mathbb{H}_2^{\mathrm{c}}(\Gamma_{40}^{\phantom{1}})+12\mathrm{X}_1^{\phantom{1}}\big)\Big)\stackrel{\mathrm{s.p.}}{=}0,
\end{equation}
where in the second transition the relations have been used
\begin{align}\label{NK-f-19}
\mathbb{H}_2^{\mathrm{c}}(\Gamma_{40}^{\phantom{1}})+12\mathrm{X}_1^{\phantom{1}}&=
6{\centering\adjincludegraphics[width = 1.8 cm, valign=c]{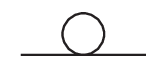}}+
12{\centering\adjincludegraphics[width = 1.8 cm, valign=c]{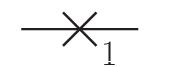}}
\\\nonumber&=
\int_{\mathbb{R}^3}\mathrm{d}^3x\,\phi^2(x)\bigg(6\Big(t_4+t_6B^2(x)/2\Big)G^\Lambda(x,x)+12\Big(m^2z_{21}+t_4z_{41}B^2(x)/2\Big)\bigg)\\\nonumber&=6
\int_{\mathbb{R}^3}\mathrm{d}^3x\,\phi^2(x)\Big(t_4+t_6B^2(x)/2\Big)PS(x,x),
\end{align}
leading to a reduction in subdivergence. Note that after the reduction of the internal singularity, the diagrams from \eqref{NK-ff-7} became convergent. This is in full agreement with the general theory, since the resulting diagrams no longer contain loops and a large ($>2$) number of identical lines.\\

\noindent\textbf{Relation 2.} Let us consider three diagrams $\{\mathrm{A}_3,\mathrm{A}_2,\mathrm{A}_1\}$ from \eqref{NK-ff-8} taking into account the corresponding coefficient from \eqref{NK-f-14}. In this case, singularities appear due to the presence of the Green's function on the diagonal, i.e. a loop, so the reduction should be done using \eqref{NK-f-19}. In this case, as a counter-term diagram, we should choose the part of the seventh term in \eqref{NK-f-15} that does not contain loops. It is easy to check that, using the loop decomposition
\begin{equation}\label{NK-f-24}
\mathbb{H}_{2}^{\mathrm{c}}(\Gamma_{30}^2\Gamma_{40}^{\phantom{1}})=\mathbb{H}_{2,0}^{\mathrm{c}}(\Gamma_{30}^2\Gamma_{40}^{\phantom{1}})+\mathbb{H}_{2,1}^{\mathrm{c}}(\Gamma_{30}^2\Gamma_{40}^{\phantom{1}}),
\end{equation}
the following relation holds
\begin{equation}\label{NK-f-24-1}
\mathbb{H}_0^{\mathrm{sc}}\big(\Gamma_{30}^2\Gamma_{40}^{\phantom{1}}\mathrm{X}_1^{\phantom{1}}\big)=
\mathbb{H}_0^{\mathrm{sc}}\big(\mathbb{H}_{2}^{\mathrm{c}}(\Gamma_{30}^2\Gamma_{40}^{\phantom{1}})\mathrm{X}_1^{\phantom{1}}\big)=\mathbb{H}_0^{\mathrm{sc}}\big(\mathbb{H}_{2,0}^{\mathrm{c}}(\Gamma_{30}^2\Gamma_{40}^{\phantom{1}})\mathrm{X}_1^{\phantom{1}}\big)+\mathbb{H}_0^{\mathrm{sc}}\big(\mathbb{H}_{2,1}^{\mathrm{c}}(\Gamma_{30}^2\Gamma_{40}^{\phantom{1}})\mathrm{X}_1^{\phantom{1}}\big).
\end{equation}
As a result, the equality is true
\begin{equation}\label{NK-f-20}
\frac{2\mathbb{H}_0^{\mathrm{sc}}\big(\mathbb{H}_{2,0}^{\mathrm{c}}(\Gamma_{30}^2\Gamma_{40}^{\phantom{1}})\mathbb{H}_2^{\mathrm{c}}(\Gamma_{40}^{\phantom{1}})\big)}{4(3!)^2(4!)^2}+
\frac{\mathbb{H}_0^{\mathrm{sc}}\big(\mathbb{H}_{2,0}^{\mathrm{c}}(\Gamma_{30}^2\Gamma_{40}^{\phantom{1}})\mathrm{X}_1^{\phantom{1}}\big)}{4(3!)^24!}\stackrel{\mathrm{s.p.}}{=}0.
%\frac{1}{2(3!)^2(4!)^2}
%\mathbb{H}_0^{\mathrm{sc}}\Big(\mathbb{H}_0^{\mathrm{sc}}(\Gamma_{30}^2\Gamma_{40}^{\phantom{1}})\big(\mathbb{H}_2^{\mathrm{c}}(\Gamma_{40}^{\phantom{1}})+12\mathrm{X}_1^{\phantom{1}}\big)\Big)
\end{equation}
Note that after reduction the diagrams no longer contain singular parts.\\

\noindent\textbf{Relation 3.} Let us further combine the diagrams $\{\mathrm{A}_5,\mathrm{A}_4\}$ from \eqref{NK-ff-8}. At the same time, we add to them the remaining part from the seventh term in \eqref{NK-f-15}, that is, the part with one loop, as well as the third term from \eqref{NK-f-16}. Thus, the combination has the form
\begin{equation}\label{NK-f-21}
\frac{\mathbb{H}_0^{\mathrm{sc}}\big(\Gamma_{30}^2\mathbb{H}_2^{\mathrm{c}}(\Gamma_{40}^{\phantom{1}})\mathbb{H}_2^{\mathrm{c}}(\Gamma_{40}^{\phantom{1}})\big)}{4(3!)^2(4!)^2}+
\frac{\mathbb{H}_0^{\mathrm{sc}}\big(\mathbb{H}_{2,1}^{\mathrm{c}}(\Gamma_{30}^2\Gamma_{40}^{\phantom{1}})\mathrm{X}_1^{\phantom{1}}\big)}{4(3!)^24!}+\frac{\mathbb{H}_0^{\mathrm{sc}}\big(\Gamma_{30}^2\mathrm{X}_1^{2}\big)}{2^4(3!)^2}.
\end{equation}
Let us replace the second term using the equality
\begin{equation}\label{NK-f-22}
\mathbb{H}_0^{\mathrm{sc}}\big(\mathbb{H}_{2,1}^{\mathrm{c}}(\Gamma_{30}^2\Gamma_{40}^{\phantom{1}})\mathrm{X}_1^{\phantom{1}}\big)=
\mathbb{H}_0^{\mathrm{sc}}\big(\Gamma_{30}^2\mathbb{H}_2^{\mathrm{c}}(\Gamma_{40}^{\phantom{1}})\mathrm{X}_1^{\phantom{1}}\big),
\end{equation}
then \eqref{NK-f-21} can be transformed using Newton's binomial formula
\begin{equation}\label{NK-f-23}
\frac{1}{4(3!)^2(4!)^2}
\mathbb{H}_0^{\mathrm{sc}}\Big(\Gamma_{30}^2\big(\mathbb{H}_2^{\mathrm{c}}(\Gamma_{40}^{\phantom{1}})+12\mathrm{X}_1^{\phantom{1}}\big)^2\Big)\stackrel{\mathrm{s.p.}}{=}0,
\end{equation}
after which all singularities in the diagrams can be reduced.\\

\noindent\textbf{Relation 4.} Next, we consider the diagrams $\mathrm{A}_{17}$ from \eqref{NK-ff-4}, $\mathrm{A}_{14}$ from \eqref{NK-ff-5}, and $\mathrm{A}_{6}$ from \eqref{NK-ff-8}. All of these contributions contain the characteristic element $\mathbb{H}_2^{\mathrm{sc}}(\Gamma_{30}^2)$, so we need to add the corresponding counter-term diagrams to them: the first term from \eqref{NK-f-16} and the part with the loop from the third term in \eqref{NK-f-15}. The resulting combination can be represented as follows
\begin{multline}\label{NK-f-25}
-\frac{\mathbb{H}_0^{\mathrm{sc}}\big(\mathbb{H}_2^{\mathrm{sc}}(\Gamma_{30}^2)\mathbb{H}_2^{\mathrm{sc}}(\Gamma_{60}^{\phantom{1}})\big)}{2(3!)^26!}+
\frac{3\mathbb{H}_0^{\mathrm{sc}}\big(\mathbb{H}_2^{\mathrm{sc}}(\Gamma_{30}^2)\mathbb{H}_{2,0}^{\mathrm{c}}(\Gamma_{30}^{\phantom{1}}\Gamma_{50}^{\phantom{1}})\big)}{(3!)^45!}+
\frac{\mathbb{H}_0^{\mathrm{sc}}\big(\mathbb{H}_2^{\mathrm{sc}}(\Gamma_{30}^2)\mathbb{H}_2^{\mathrm{sc}}(\Gamma_{40}^2)\big)}{4(3!)^2(4!)^2}-\\-
\frac{\mathbb{H}_0^{\mathrm{sc}}\big(\mathbb{H}_2^{\mathrm{sc}}(\Gamma_{30}^2)\mathrm{X}_2^{\phantom{1}}\big)}{4(3!)^2}-
\frac{\mathbb{H}_0^{\mathrm{sc}}\big(\mathbb{H}_2^{\mathrm{sc}}(\Gamma_{30}^2)\mathbb{H}_2^{\mathrm{c}}(\Gamma_{41}^{\phantom{1}})\big)}{2(3!)^24!}=\frac{1}{8(3!)^2}
\mathbb{H}_0^{\mathrm{sc}}\big(\mathbb{H}_2^{\mathrm{sc}}(\Gamma_{30}^2)\widetilde\Gamma_2^{\phantom{1}}\big),
\end{multline}
where
\begin{equation}\label{NK-f-26}
\widetilde\Gamma_2^{\phantom{1}}=-\frac{4}{6!}\mathbb{H}_2^{\mathrm{sc}}(\Gamma_{60}^{\phantom{1}})
+\frac{4}{6!}\mathbb{H}_{2,0}^{\mathrm{c}}(\Gamma_{30}^{\phantom{1}}\Gamma_{50}^{\phantom{1}})
+\frac{2}{(4!)^2}\mathbb{H}_2^{\mathrm{sc}}(\Gamma_{40}^2)-2\mathrm{X}_2^{\phantom{1}}
-\frac{1}{6}\mathbb{H}_2^{\mathrm{c}}(\Gamma_{41}^{\phantom{1}}).
\end{equation}
The last combination can also be represented in the diagrammatic form
\begin{equation}\label{NK-f-28}-\frac{1}{4}
{\centering\adjincludegraphics[width = 1.5 cm, valign=c]{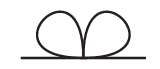}}+\frac{1}{3}
{\centering\adjincludegraphics[width = 1.3 cm, valign=c]{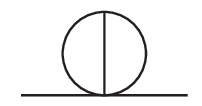}}+\frac{1}{3}
{\centering\adjincludegraphics[width = 1.3 cm, valign=c]{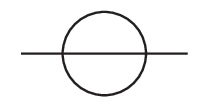}}-2
{\centering\adjincludegraphics[width = 1.5 cm, valign=c]{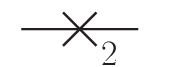}}-
{\centering\adjincludegraphics[width = 1.3 cm, valign=c]{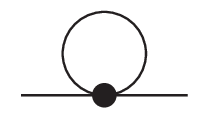}}.
\end{equation}
Further, it can be verified by direct calculation that the density does not contain singular components
\begin{multline}\label{NK-f-27}
-\frac{t_6}{4}\big(G^\Lambda(x,x)\big)^2+\frac{1}{3}\frac{L}{16\pi^2}
\bigg(t_4B(x)+\frac{t_6}{6}B^3(x)\bigg)t_6B(x)+\frac{1}{3}\frac{L}{16\pi^2}
\bigg(t_4+\frac{t_6}{2}B^2(x)\bigg)^2-\\-2\bigg(
\frac{t_6}{8}\alpha^2\Lambda^2+\frac{Lt_4^2}{96\pi^2}+
\frac{Lt_4t_6B^2(x)}{48\pi^2}+\frac{5Lt_6^2B^4(x)}{48\pi^24!}\bigg)+t_4\frac{\alpha\Lambda t_6}{2t_4}G^\Lambda(x,x)\stackrel{\mathrm{s.p.}}{=}0,
\end{multline}
where for the second and third diagrams an asymptotic expansion of the form has been used
\begin{equation}\label{NK-f-30}
\int_{\mathrm{B}_{1/\sigma}}\mathrm{d}^3x\,\Big(G^\Lambda(x+y,y)\Big)^3\stackrel{\mathrm{s.p.}}{=}\frac{L}{16\pi^2}.
\end{equation}
Finally, we obtain that the combination
\begin{equation}\label{NK-f-29}
\mathbb{H}_0^{\mathrm{sc}}\big(\mathbb{H}_2^{\mathrm{sc}}(\Gamma_{30}^2)\widetilde\Gamma_2^{\phantom{1}}\big)\stackrel{\mathrm{s.p.}}{=}0
\end{equation}
does not contain singular parts and the linear combination \eqref{NK-f-25} is finite.\\

\noindent\textbf{Relation 5.} Consider the diagrams $\mathrm{A}_{25}$ from \eqref{NK-ff-1}, $\mathrm{A}_{23}$ from \eqref{NK-ff-2}, and $\mathrm{A}_{13}$ from \eqref{NK-ff-6}. They contain a common subdiagram of the form $\mathbb{H}_2^{\mathrm{c}}(\Gamma_{40}^{\phantom{1}})$, so we add to them the corresponding counter-term diagrams: the second term from \eqref{NK-f-16} and the loop part of the fourth term in \eqref{NK-f-15}. The resulting combination is
\begin{multline}\label{NK-f-31}
\frac{\mathbb{H}_0^{\mathrm{sc}}\big(\mathbb{H}_2^{\mathrm{c}}(\Gamma_{40}^{\phantom{1}})\mathbb{H}_2^{\mathrm{sc}}(\Gamma_{60}^{\phantom{1}})\big)}{4!6!}-
\frac{\mathbb{H}_0^{\mathrm{sc}}\big(\mathbb{H}_2^{\mathrm{c}}(\Gamma_{40}^{\phantom{1}})\mathbb{H}_{2,0}^{\mathrm{c}}(\Gamma_{30}^{\phantom{1}}\Gamma_{50}^{\phantom{1}})\big)}{3!4!5!}-
\frac{3\mathbb{H}_0^{\mathrm{sc}}\big(\mathbb{H}_2^{\mathrm{c}}(\Gamma_{40}^{\phantom{1}})\mathbb{H}_2^{\mathrm{sc}}(\Gamma_{40}^2)\big)}{3!(4!)^3}+\\+
\frac{\mathbb{H}_0^{\mathrm{sc}}\big(\mathbb{H}_2^{\mathrm{c}}(\Gamma_{40}^{\phantom{1}})\mathrm{X}_2^{\phantom{1}}\big)}{2(4!)}+
\frac{\mathbb{H}_0^{\mathrm{sc}}\big(\mathbb{H}_2^{\mathrm{c}}(\Gamma_{40}^{\phantom{1}})\mathbb{H}_2^{\mathrm{c}}(\Gamma_{41}^{\phantom{1}})\big)}{(4!)^2}=-\frac{1}{4(4!)}
\mathbb{H}_0^{\mathrm{sc}}\big(\mathbb{H}_2^{\mathrm{c}}(\Gamma_{40}^{\phantom{1}})\widetilde\Gamma_2^{\phantom{1}}\big).
\end{multline}
It is clear that the vertex $\widetilde\Gamma_2^{\phantom{1}}$ does not contain singularities due to \eqref{NK-f-27}. However, the first factor is singular, so it is necessary to choose a set of counter-term diagrams from \eqref{NK-f-15}, \eqref{NK-f-16}, and \eqref{NK-f-17} of the form
\begin{multline}\label{NK-f-32}
\frac{\mathbb{H}_0^{\mathrm{sc}}\big(\mathrm{X}_1^{\phantom{1}}\mathbb{H}_2^{\mathrm{sc}}(\Gamma_{60}^{\phantom{1}})\big)}{2(6!)}-
\frac{\mathbb{H}_0^{\mathrm{sc}}\big(\mathrm{X}_1^{\phantom{1}}\mathbb{H}_{2,0}^{\mathrm{c}}(\Gamma_{30}^{\phantom{1}}\Gamma_{50}^{\phantom{1}})\big)}{2(3!5!)}-
\frac{\mathbb{H}_0^{\mathrm{sc}}\big(\mathrm{X}_1^{\phantom{1}}\mathbb{H}_2^{\mathrm{sc}}(\Gamma_{40}^2)\big)}{4(4!)^2}+\\+
\frac{\mathbb{H}_0^{\mathrm{sc}}\big(\mathrm{X}_1^{\phantom{1}}\mathrm{X}_2^{\phantom{1}}\big)}{4}+
\frac{\mathbb{H}_0^{\mathrm{sc}}\big(\mathrm{X}_1^{\phantom{1}}\mathbb{H}_2^{\mathrm{c}}(\Gamma_{41}^{\phantom{1}})\big)}{2(4!)}=-\frac{1}{8}
\mathbb{H}_0^{\mathrm{sc}}\big(\mathrm{X}_1^{\phantom{1}}\widetilde\Gamma_2^{\phantom{1}}\big).
\end{multline}
It is clear that it has a suitable form, such that the combinations \eqref{NK-f-31} and \eqref{NK-f-32} in sum
\begin{equation}\label{NK-f-33}
-\frac{1}{4(4!)}
\mathbb{H}_0^{\mathrm{sc}}\Big(\big(\mathbb{H}_2^{\mathrm{c}}(\Gamma_{40}^{\phantom{1}})+12\mathrm{X}_1^{\phantom{1}}\big)\widetilde\Gamma_2^{\phantom{1}}\Big)\stackrel{\mathrm{s.p.}}{=}0
\end{equation}
do not contain singular contributions due to \eqref{NK-f-19}.\\

\noindent\textbf{Relation 6.} Let us study the diagram $\mathrm{A}_{18}$ from \eqref{NK-ff-4}. Its singularity is one Green's function on the diagonal, that is, a loop formed from the vertex $\Gamma_{60}^{\phantom{1}}$. It is clear that this type of singularity can be removed by the first coefficient contained in $\Gamma_{41}^{\phantom{1}}$. Therefore, after adding the corresponding part from the third term in \eqref{NK-f-15}, we obtain
\begin{multline}\label{NK-f-34}
-\frac{2\mathbb{H}_0^{\mathrm{sc}}\big(\Gamma_{30}^{\phantom{1}}
\mathbb{H}_3^{\mathrm{sc}}(\Gamma_{30}^{\phantom{1}}
\mathbb{H}_4^{\mathrm{sc}}(\Gamma_{60}^{\phantom{1}}))\big)}{2(3!)^26!}
-\frac{2\mathbb{H}_0^{\mathrm{sc}}\big(\Gamma_{30}^{\phantom{1}}
	\mathbb{H}_3^{\mathrm{sc}}(\Gamma_{30}^{\phantom{1}}
	\Gamma_{41}^{\phantom{1}})\big)}{2(3!)^24!}=\\=
-\frac{1}{(3!)^26!}
\mathbb{H}_0^{\mathrm{sc}}\Big(\Gamma_{30}^{\phantom{1}}
\mathbb{H}_3^{\mathrm{sc}}\big(\Gamma_{30}^{\phantom{1}}\big[
\mathbb{H}_4^{\mathrm{sc}}(\Gamma_{60}^{\phantom{1}})+30\Gamma_{41}^{\phantom{1}}\big]\big)\Big)
\stackrel{\mathrm{s.p.}}{=}0.
\end{multline}
Thus, we again obtain the finite diagram after removing the internal singularity.\\

\noindent\textbf{Relation 7.} Consider the diagram $\mathrm{A}_{26}$ from \eqref{NK-ff-1}. ​​The distinctive feature is the presence of two Green's functions on the diagonal, which are obtained due to the presence of two vertices $\mathbb{H}_2^{\mathrm{c}}(\Gamma_{40}^{\phantom{1}})$. Adding a number of suitable counter-term diagrams from the eighth term in \eqref{NK-f-15} and the fourth term in \eqref{NK-f-16}, we obtain the following linear combination
\begin{equation}\label{NK-f-35}
-\frac{3
\mathbb{H}_0^{\mathrm{sc}}\big(
\mathbb{H}_{2,1}^{\mathrm{c}}\big(\Gamma_{40}^{\phantom{1}}
\mathbb{H}_2^{\mathrm{c}}(\Gamma_{40}^{\phantom{1}})\big)\mathbb{H}_2^{\mathrm{c}}(\Gamma_{40}^{\phantom{1}})\big)}{3!(4!)^3}
-\frac{2
	\mathbb{H}_0^{\mathrm{sc}}\big(
	\mathbb{H}_{2,0}^{\mathrm{c}}\big(\Gamma_{40}^{\phantom{1}}
	\mathrm{X}_1^{\phantom{1}}\big)\mathbb{H}_2^{\mathrm{c}}(\Gamma_{40}^{\phantom{1}})\big)}{4(4!)^2}
-\frac{
	\mathbb{H}_0^{\mathrm{sc}}\big(
	\mathbb{H}_{2,0}^{\mathrm{c}}\big(\Gamma_{40}^{\phantom{1}}
	\mathrm{X}_1^{\phantom{1}}\big)\mathrm{X}_1^{\phantom{1}}\big)}{8(4!)},
\end{equation}
which after applying the equality
\begin{equation*}
\mathbb{H}_0^{\mathrm{sc}}\Big(
\mathbb{H}_{2,1}^{\mathrm{c}}\big(\Gamma_{40}^{\phantom{1}}
	\mathbb{H}_2^{\mathrm{c}}(\Gamma_{40}^{\phantom{1}})\big)\mathrm{X}_1^{\phantom{1}}\Big)=
\mathbb{H}_0^{\mathrm{sc}}\Big(
	\mathbb{H}_{2,0}^{\mathrm{c}}\big(\Gamma_{40}^{\phantom{1}}
	\mathrm{X}_1^{\phantom{1}}\big)\mathbb{H}_2^{\mathrm{c}}(\Gamma_{40}^{\phantom{1}})\Big)
\end{equation*}
can be represented as
\begin{equation}\label{NK-f-36}
-\frac{1}{2(4!)^3}
\mathbb{H}_0^{\mathrm{sc}}\Big(\Big[
\mathbb{H}_{2,1}^{\mathrm{c}}\big(\Gamma_{40}^{\phantom{1}}
\mathbb{H}_2^{\mathrm{c}}(\Gamma_{40}^{\phantom{1}})\big)+12\mathbb{H}_{2,0}^{\mathrm{c}}\big(\Gamma_{40}^{\phantom{1}}
\mathrm{X}_1^{\phantom{1}}\big)\Big]
\Big[\mathbb{H}_2^{\mathrm{c}}(\Gamma_{40}^{\phantom{1}})+12\mathrm{X}_1^{\phantom{1}}\Big]\Big)
\stackrel{\mathrm{s.p.}}{=}0.
\end{equation}
The last equality follows after applying the relation \eqref{NK-f-19} to both factors.\\

\noindent\textbf{Relation 8.} Let us consider the diagram $\mathrm{A}_{27}$ from \eqref{NK-ff-1}. ​​The Green's function appears three times on the diagonal, so the combination should be assembled taking into account  equality \eqref{NK-f-19}. Let us add to the diagram the corresponding counter-term diagrams from the eighth term in \eqref{NK-f-15}, the fourth term in \eqref{NK-f-16}, and the first in \eqref{NK-f-17}, then we get the combination
\begin{equation}\label{NK-f-37}
-\frac{\mathbb{H}_0^{\mathrm{sc}}\big(\big(\mathbb{H}_2^{\mathrm{c}}(\Gamma_{40}^{\phantom{1}})\big)^3\big)
}{3!(4!)^3}
-\frac{\mathbb{H}_0^{\mathrm{sc}}\big(\big(\mathbb{H}_2^{\mathrm{c}}(\Gamma_{40}^{\phantom{1}})\big)^2\mathrm{X}_1^{\phantom{1}}\big)
}{4(4!)^2}
-\frac{\mathbb{H}_0^{\mathrm{sc}}\big(\mathbb{H}_2^{\mathrm{c}}(\Gamma_{40}^{\phantom{1}})\mathrm{X}_1^{2}\big)
}{8(4!)}
-\frac{\mathbb{H}_0^{\mathrm{sc}}\big(\mathrm{X}_1^{3}\big)
}{8(3!)},
\end{equation}
which after reduction of similars is transformed into
\begin{equation}\label{NK-f-38}
-\frac{1}{3!(4!)^3}
\mathbb{H}_0^{\mathrm{sc}}\Big(\big(\mathbb{H}_2^{\mathrm{c}}(\Gamma_{40}^{\phantom{1}})+12\mathrm{X}_1^{\phantom{1}}\big)^3\Big)\stackrel{\mathrm{s.p.}}{=}0.
\end{equation}
Thus, the result has no singular components.\\

\noindent\textbf{Relation 9.} Let us turn to $\mathrm{A}_{16}$ and $\mathrm{A}_{15}$ from \eqref{NK-ff-5}. Their distinctive feature is the presence of the Green's function on the diagonal in the form of the vertex $\mathbb{H}_3^{\mathrm{c}}(\Gamma_{50}^{\phantom{1}})$. It is clear that such a singularity should be cancelled by the counter-vertex $\Gamma_{31}^{\phantom{1}}$. By adding the first term from \eqref{NK-f-15}, we can verify the equality
\begin{equation}\label{NK-f-39}
\frac{\mathbb{H}_0^{\mathrm{sc}}
\big(\Gamma_{30}^3\mathbb{H}_3^{\mathrm{c}}(\Gamma_{50}^{\phantom{1}})\big)
}{(3!)^45!}+\frac{\mathbb{H}_0^{\mathrm{sc}}
\big(\Gamma_{30}^3\Gamma_{31}^{\phantom{1}}\big)
}{(3!)^5}=\frac{1}{(3!)^45!}
\mathbb{H}_0^{\mathrm{sc}}
\Big(\Gamma_{30}^3
\big[\mathbb{H}_3^{\mathrm{c}}(\Gamma_{50}^{\phantom{1}})+20\Gamma_{31}^{\phantom{1}}\big]\Big)
\stackrel{\mathrm{s.p.}}{=}0.
\end{equation}

\noindent\textbf{Relation 10.} Similar reasoning is also true for the diagram $\mathrm{A}_{12}$ from \eqref{NK-ff-6}. Adding the part from the second term in \eqref{NK-f-15}, we obtain the following equality
\begin{equation}\label{NK-f-40}
-\frac{\mathbb{H}_0^{\mathrm{sc}}
	\big(\mathbb{H}_3^{\mathrm{sc}}(\Gamma_{30}^{\phantom{1}}\Gamma_{40}^{\phantom{1}})\mathbb{H}_3^{\mathrm{c}}(\Gamma_{50}^{\phantom{1}})\big)
}{3!4!5!}-\frac{\mathbb{H}_0^{\mathrm{sc}}
\big(\mathbb{H}_3^{\mathrm{sc}}(\Gamma_{30}^{\phantom{1}}\Gamma_{40}^{\phantom{1}})\Gamma_{31}^{\phantom{1}}\big)
}{(3!)^24!}\stackrel{\mathrm{s.p.}}{=}0.
\end{equation}

\noindent\textbf{Relation 11.} Next, consider $\mathrm{A}_{11}$ from \eqref{NK-ff-6}. This diagram contains two Green's functions on the diagonal, which follow from the vertices $\mathbb{H}_2^{\mathrm{c}}(\Gamma_{40}^{\phantom{1}})$ and $\mathbb{H}_3^{\mathrm{c}}(\Gamma_{50}^{\phantom{1}})$. Therefore, such singularities must be cancelled by the counter-vertices $\mathrm{X}_1^{\phantom{1}}$ and $\Gamma_{31}^{\phantom{1}}$. Indeed, adding the remainder of the second term in \eqref{NK-f-15}, the remainder of the ninth term in \eqref{NK-f-15}, and the eighth term in \eqref{NK-f-16}, we obtain the combination
\begin{equation}\label{NK-f-41}
-\frac{\mathbb{H}_0^{\mathrm{sc}}
	\big(\Gamma_{30}^{\phantom{1}}\mathbb{H}_2^{\mathrm{c}}(\Gamma_{40}^{\phantom{1}})\mathbb{H}_3^{\mathrm{c}}(\Gamma_{50}^{\phantom{1}})\big)
}{3!4!5!}
-\frac{\mathbb{H}_0^{\mathrm{sc}}
	\big(\Gamma_{30}^{\phantom{1}}\mathbb{H}_2^{\mathrm{c}}(\Gamma_{40}^{\phantom{1}})\Gamma_{31}^{\phantom{1}}\big)
}{(3!)^24!}
-\frac{\mathbb{H}_0^{\mathrm{sc}}
	\big(\Gamma_{30}^{\phantom{1}}\mathrm{X}_1^{\phantom{1}}\mathbb{H}_3^{\mathrm{c}}(\Gamma_{50}^{\phantom{1}})\big)
}{2(3!5!)}
-\frac{\mathbb{H}_0^{\mathrm{sc}}
	\big(\Gamma_{30}^{\phantom{1}}\mathrm{X}_1^{\phantom{1}}\Gamma_{31}^{\phantom{1}}\big)
}{2(3!)^2},
\end{equation}
which is transformed into the form
\begin{equation}\label{NK-f-42}
-\frac{1}{3!4!5!}
\mathbb{H}_0^{\mathrm{sc}}
\Big(\Gamma_{30}^{\phantom{1}}
\big[\mathbb{H}_2^{\mathrm{c}}(\Gamma_{40}^{\phantom{1}})+12\mathrm{X}_1^{\phantom{1}}\big]
\big[\mathbb{H}_3^{\mathrm{c}}(\Gamma_{50}^{\phantom{1}})+20\Gamma_{31}^{\phantom{1}}\big]\Big)
\stackrel{\mathrm{s.p.}}{=}0.
\end{equation}
Thus, the result has no singular parts.\\

\noindent\textbf{Relation 12.} Consider $\mathrm{A}_{22}$ from \eqref{NK-ff-2} together with the corresponding fourth term from \eqref{NK-f-15}. The linear combination is
\begin{equation}\label{NK-f-43}
\frac{\mathbb{H}_{0,1}^{\mathrm{sc}}
	\big(\Gamma_{40}^{\phantom{1}}\mathbb{H}_4^{\mathrm{c}}(\Gamma_{60}^{\phantom{1}})\big)
}{4!6!}+\frac{\mathbb{H}_{0,0}^{\mathrm{sc}}
\big(\Gamma_{40}^{\phantom{1}}\Gamma_{41}^{\phantom{1}}\big)
}{(4!)^2}=\frac{1}{48}\bigg(
{\centering\adjincludegraphics[width = 1.8 cm, valign=c]{NK-f-22.eps}}+2
{\centering\adjincludegraphics[width = 1.6 cm, valign=c]{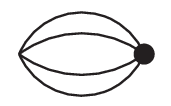}}\bigg).
\end{equation}
The analytical expression is as follows
\begin{equation}\label{NK-f-44}
\frac{1}{48}\int_{\mathbb{R}^3}\mathrm{d}^3y\int_{\mathbb{R}^3}\mathrm{d}^3x\,
\bigg(t_4+\frac{t_6B^2(y)}{2}\bigg)\Big(G^\Lambda(y,x)\Big)^4\Big(t_6\alpha\Lambda+t_6PS(x,x)+2t_4z_{41}\Big).
\end{equation}
Taking into account the equality $t_6\alpha\Lambda+2t_4z_{41}=0$, we immediately obtain the singular component
\begin{equation}\label{NK-f-45}
\frac{\alpha_1(\mathbf{f})\Lambda}{48}\int_{\mathbb{R}^3}\mathrm{d}^3x\,
\bigg(t_4+\frac{t_6B^2(x)}{2}\bigg)t_6PS(x,x)+\frac{L}{12(16\pi^2)}\int_{\mathbb{R}^3}\mathrm{d}^3x\,
\bigg(t_4+\frac{t_6B^2(x)}{2}\bigg)t_6PS^2(x,x),
\end{equation}
where an auxiliary number  has been entered
\begin{equation}\label{NK-f-46}
\alpha_1(\mathbf{f})=\int_{\mathbb{R}^3}\mathrm{d}^3y\,\Big(R_0^1(y)\Big)^4.
\end{equation}
Note that the entire singular part depends on the nonlocal function $PS$.\\

\noindent\textbf{Relation 13.}The following diagram $\mathrm{A}_{20}$ from \eqref{NK-ff-3} contains two vertices $\mathbb{H}_3^{\mathrm{c}}(\Gamma_{50}^{\phantom{1}})$, so the subtrahends should be the fifth term from \eqref{NK-f-15} and the seventh term from \eqref{NK-f-16}. As a result, we obtain
\begin{equation}\label{NK-f-47}
\frac{\mathbb{H}_0^{\mathrm{sc}}
	\big(\big(\mathbb{H}_3^{\mathrm{c}}(\Gamma_{50}^{\phantom{1}})\big)^2\big)
}{2(5!)^2}+
\frac{\mathbb{H}_0^{\mathrm{sc}}
	\big(\Gamma_{31}^{\phantom{1}}\mathbb{H}_3^{\mathrm{c}}(\Gamma_{50}^{\phantom{1}})\big)
}{3!5!}+
\frac{\mathbb{H}_0^{\mathrm{sc}}
	\big(\big(\Gamma_{31}^{\phantom{1}}\big)^2\big)
}{2(3!)^2}=\frac{1}{2(5!)^2}
\mathbb{H}_0^{\mathrm{sc}}
	\Big(\big[\mathbb{H}_3^{\mathrm{c}}(\Gamma_{50}^{\phantom{1}})+20\Gamma_{31}^{\phantom{1}}\big]^2\Big).
\end{equation}
It is clear that it contains a singular component
\begin{equation}\label{NK-f-48}
\frac{1}{48}\bigg(
{\centering\adjincludegraphics[width = 2.3 cm, valign=c]{NK-f-20.eps}}+2
{\centering\adjincludegraphics[width = 2.1 cm, valign=c]{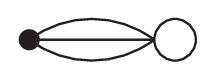}}+4
{\centering\adjincludegraphics[width = 1.8 cm, valign=c]{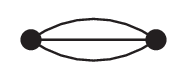}}
\bigg)\stackrel{\mathrm{s.p.}}{=}\frac{Lt_6^2}{48(16\pi^2)}
\int_{\mathbb{R}^3}\mathrm{d}^3x\,PS^2(x,x)B^2(x),
\end{equation}
which has a nonlocal character (depends on the function $PS$).\\

\noindent\textbf{Relation 14.} Next, we study the linear combination of the diagrams $\mathrm{A}_{19}$ from \eqref{NK-ff-4} and $\mathrm{A}_{10}$ from \eqref{NK-ff-6}, as well as the fifth term from \eqref{NK-f-16}
\begin{equation}\label{NK-f-49}
-\frac{\mathbb{H}_{0,0}^{\mathrm{sc}}
	\big(\Gamma_{30}^{2}\Gamma_{60}^{\phantom{1}}\big)
}{2(3!)^26!}
-\frac{\mathbb{H}_{0,0}^{\mathrm{sc}}
	\big(\Gamma_{30}^{\phantom{1}}\Gamma_{40}^{\phantom{1}}\Gamma_{50}^{\phantom{1}}\big)
}{3!4!5!}+\frac{\mathbb{H}_{0,0}^{\mathrm{sc}}
\big(\Gamma_{30}^{\phantom{1}}\Gamma_{32}^{\phantom{1}}\big)
}{(3!)^2},
\end{equation}
which in diagrammatic language takes the form
\begin{equation}\label{NK-f-50}
-\frac{1}{72}
{\centering\adjincludegraphics[width = 1.8 cm, valign=c]{NK-f-19.eps}}
-\frac{1}{12}
{\centering\adjincludegraphics[width = 1.4 cm, valign=c]{NK-f-10.eps}}
+\frac{1}{6}
{\centering\adjincludegraphics[width = 1.2 cm, valign=c]{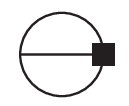}}.
\end{equation}
Let us find asymptotic expansions for each of the terms separately. The first diagram can be rewritten analytically
\begin{equation}\label{NK-f-51}
{\centering\adjincludegraphics[width = 1.8 cm, valign=c]{NK-f-19.eps}}=
\int_{\mathbb{R}^{3\times3}}\mathrm{d}^3x\mathrm{d}^3y\mathrm{d}^3z\,
r(x)\Big(G^\Lambda(x,y)\Big)^3t_6
\Big(G^\Lambda(y,z)\Big)^3r(z),
\end{equation}
where for convenience the function was introduced
\begin{equation}\label{NK-f-52}
r(x)=\bigg(t_4B(x)+\frac{t_6B^3(x)}{3!}\bigg).
\end{equation}
Next, it is convenient to use addition and subtraction and represent each side multiplier as follows
\begin{equation}\label{NK-f-53}
r(x)\Big(G^\Lambda(x,y)\Big)^3=
r(x)\Big(G^\Lambda(x,y)\Big)^3\pm
r(y)\Big(\widetilde{R}_0^\Lambda(x-y)\Big)^3,
\end{equation}
\begin{equation}\label{NK-f-54}
r(z)\Big(G^\Lambda(y,z)\Big)^3=
r(z)\Big(G^\Lambda(y,z)\Big)^3\pm
r(y)\Big(\widetilde{R}_0^\Lambda(z-y)\Big)^3,
\end{equation}
where the cut function is defined as
\begin{equation}\label{NK-f-55}
\widetilde{R}_0^\Lambda(x)=R_0^\Lambda(x)\,\chi\big(1/\Lambda<|x|<1/\sigma\big).
\end{equation}
Then, discarding the finite parts and using the integral equality
\begin{equation}\label{NK-f-56}
\int_{\mathbb{R}^{3}}\mathrm{d}^3x\,\Big(\widetilde{R}_0^\Lambda(x)\Big)^3=\frac{L}{16\pi^2},
\end{equation}
we arrive at the following expression for the first diagram
\begin{equation}\label{NK-f-57}
{\centering\adjincludegraphics[width = 1.8 cm, valign=c]{NK-f-19.eps}}\stackrel{\mathrm{s.p.}}{=}
\frac{2L}{16\pi^2}
\int_{\mathbb{R}^{3\times2}}\mathrm{d}^3x\mathrm{d}^3y\,
r(x)\Big(G^\Lambda(x,y)\Big)^3t_6r(y)-\bigg(\frac{L}{16\pi^2}\bigg)^2
\int_{\mathbb{R}^{3}}\mathrm{d}^3x\,t_6r^2(x).
\end{equation}
Let us move on to the next diagram and use addition and subtraction again
\begin{equation}\label{NK-f-58}
{\centering\adjincludegraphics[width = 1.4 cm, valign=c]{NK-f-10.eps}}\pm\mathrm{I}_1(\Lambda)
\int_{\mathbb{R}^{3}}\mathrm{d}^3x\,r(x)t_6B(x)\bigg(t_4+\frac{t_6B^2(x)}{2}\bigg),
\end{equation}
where
\begin{equation}\label{NK-f-59}
\mathrm{I}_1(\Lambda)%&\,
=\frac{L}{16\pi^2}\int_{\mathrm{B}_{1/\sigma}}\mathrm{d}^3y\,\Big(R_0^\Lambda(y)\Big)^3-
\int_{\mathrm{B}_{1/\sigma}\times\mathrm{B}_{1/\sigma}}\mathrm{d}^3y\mathrm{d}^3z\,
\Big(R_0^\Lambda(y)\Big)^2R_0^\Lambda(y-z)\Big(R_0^\Lambda(z)\Big)^3
%\\\nonumber&\stackrel{\mathrm{s.p.}}{=}
%\frac{L}{16\pi^2}\int_{\mathrm{B}_{1/\sigma}}\mathrm{d}^3y\,\Big(R_0^\Lambda(y)\Big)^3-
%\int_{\mathrm{B}_{1/\sigma}\times\mathrm{B}_{1/\sigma}}\mathrm{d}^3y\mathrm{d}^3z\,
%\Big(R_0^{\phantom{1}}(y)\Big)^2R_0^{\phantom{1}}(y-z)\Big(R_0^\Lambda(z)\Big)^3
%\\\nonumber&\,=
%\frac{L}{16\pi^2}\int_{\mathrm{B}_{1/\sigma}}\mathrm{d}^3y\,\Big(R_0^\Lambda(y)\Big)^3-
%\int_{\mathrm{B}_{1/\sigma}}\mathrm{d}^3z\,
%\Big(R_0^\Lambda(y)\Big)^3\cdot\frac{1-\ln(|y|\sigma)}{16\pi^2}\stackrel{\mathrm{s.p.}}{=}
\stackrel{\mathrm{s.p.}}{=}-\frac{L}{(4\pi)^4}+\frac{L^2}{2(4\pi)^4}.
\end{equation}
Here we used the transition $$\Big(R_0^\Lambda(y)\Big)^2R_0^\Lambda(y-z) \stackrel{\mathrm{s.p.}}{\to} \Big(R_0^{\phantom{1}}(y)\Big)^2R_0^{\phantom{1}}(y-z) \stackrel{\int}{\to} \frac{1-\ln(|y|\sigma)}{16\pi^2},$$
and also the formula, see \cite{Ivanov-2022},
\begin{equation*}
\int_{|\hat{y}|=1}\mathrm{d}^2\sigma(\hat{y})\,R_0(x+r\hat{y})=\frac{1}{4\pi}
	\begin{cases}
		\,\,\,r^{-1}, & |x|\leqslant r;\\
		|x|^{-1},& |x|>r.
	\end{cases}
\end{equation*}
Then, by contracting the subdiagram with three lines, the singular part of formula \eqref{NK-f-58}  becomes
\begin{equation*}
\frac{L}{16\pi^2}
\int_{\mathbb{R}^{3\times2}}\mathrm{d}^3x\mathrm{d}^3y\,
r(x)\Big(G^\Lambda(x,y)\Big)^3\bigg(t_4t_6B(y)+\frac{t_6^2B^3(y)}{2}\bigg)
-\mathrm{I}_1(\Lambda)
\int_{\mathbb{R}^{3}}\mathrm{d}^3x\,r(x)t_6B(x)\bigg(t_4+\frac{t_6B(x)}{2}\bigg).
\end{equation*}
The last diagram is instantly written out, taking into account the type of the counter-vertex $\Gamma_{42}$
\begin{equation}\label{NK-f-61}
{\centering\adjincludegraphics[width = 1.2 cm, valign=c]{NK-rr-4.eps}}=
\int_{\mathbb{R}^{3\times2}}\mathrm{d}^3x\mathrm{d}^3y\,
r(x)\Big(G^\Lambda(x,y)\Big)^3\bigg(\frac{Lt_4t_6B(y)}{24\pi^2}+\frac{5Lt_6^2B^3(y)}{48\pi^23!}\bigg).
\end{equation}
Finally, returning to the linear combination \eqref{NK-f-50}, we arrive at the expression
\begin{align}
\eqref{NK-f-49}&\stackrel{\mathrm{s.p.}}{=}
\frac{1}{72}\bigg(\frac{L}{16\pi^2}\bigg)^2
\int_{\mathbb{R}^{3}}\mathrm{d}^3x\,t_6r^2(x)+\frac{\mathrm{I}_1(\Lambda)}{12}
\int_{\mathbb{R}^{3}}\mathrm{d}^3x\,r(x)t_6B(x)\bigg(t_4+\frac{t_6B(x)}{2}\bigg)
\label{NK-f-62}
\\\nonumber&\,=
\mathrm{V}_{0,2}\bigg(
\frac{L^2t_4^2t_6^{\phantom{1}}}{3^22(4\pi)^4}-\frac{Lt_4^2t_6^{\phantom{1}}}{12(4\pi)^4}
\bigg)+
\mathrm{V}_{0,4}\bigg(
\frac{7L^2t_4^{\phantom{1}}t_6^2}{2^33^3(4\pi)^4}-\frac{Lt_4^{\phantom{1}}t_6^2}{3^22(4\pi)^4}\bigg)+
\mathrm{V}_{0,6}\bigg(\frac{5L^2t_6^3}{2^43^4(4\pi)^4}-\frac{Lt_6^3}{2^43^2(4\pi)^4}\bigg).
\end{align}

\noindent\textbf{Relation 15.} Next, we study the remaining two counter-term diagrams, the sixth term from \eqref{NK-f-16}, and the third term from \eqref{NK-f-17}. They have the form
\begin{equation}\label{NK-f-63}
-\frac{\mathbb{H}_0^{\mathrm{sc}}(\Gamma_{42}^{\phantom{1}})}{4!}-
\frac{\mathbb{H}_0^{\mathrm{sc}}(\mathrm{X}_3^{\phantom{1}})}{2}=-\frac{1}{8}
{\centering\adjincludegraphics[width = 1.7 cm, valign=c]{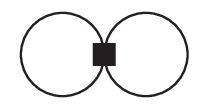}}-\frac{1}{2}
{\centering\adjincludegraphics[width = 1.5 cm, valign=c]{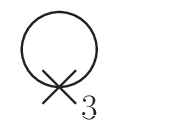}}.
\end{equation}
Using the Green's function on the diagonal, the explicit form of singularities is easily written out. We split it into two parts: one depending on the nonlocal component $PS$
\begin{multline}\label{NK-f-64}
-\frac{1}{8}\int_{\mathbb{R}^{3}}\mathrm{d}^3x\,PS^2(x,x)\bigg(\frac{Lt_4t_6}{24\pi^2}+\frac{5Lt_6^2B^2(x)}{96\pi^2}\bigg)-\frac{\alpha\Lambda}{8}\int_{\mathbb{R}^{3}}\mathrm{d}^3x\,PS(x,x)\bigg(\frac{ Lt_4t_6}{12\pi^2}+\frac{5Lt_6^2B^2(x)}{48\pi^2}\bigg)-\\
-\frac{1}{2}\int_{\mathbb{R}^{3}}\mathrm{d}^3x\,PS(x,x)\bigg(-\frac{\alpha\Lambda Lt_4t_6}{48\pi^2}+\frac{ \alpha_1(\mathbf{f})\Lambda t_4t_6}{24}-\frac{5\alpha\Lambda Lt_6^2B^2(x)}{192\pi^2}+\frac{\alpha_1(\mathbf{f})\Lambda t_6^2B^2(x)}{16}\bigg),
\end{multline}
and the part depending only on the background field $B$
\begin{equation}\label{NK-f-65}
\int_{\mathbb{R}^{3}}\mathrm{d}^3x\,B^2(x)\bigg(-\frac{\alpha^2\Lambda^2}{16}\frac{5Lt_6^2}{48\pi^2}+\frac{\alpha^2\Lambda^2}{4}\frac{5Lt_6^2}{96\pi^2}-\frac{\alpha_1(\mathbf{f})\alpha\Lambda^2}{4}\frac{t_6^2}{8}\bigg)
=
\mathrm{V}_{0,2}\bigg(
\frac{5L\alpha^2\Lambda^2t_6^2}{2^43(4\pi)^2}
-\frac{\alpha_1(\mathbf{f})\alpha\Lambda^2t_6^2}{2^5}\bigg).
\end{equation}

\noindent\textbf{Relation 16.} The answer for the diagram  $\mathrm{A}_{21}$ from \eqref{NK-ff-3} 
\begin{equation}\label{NK-f-66}
\frac{\mathbb{H}_0^{\mathrm{sc}}(\Gamma_{50}^2)}{2(5!)^2}=\frac{1}{240}{\centering\adjincludegraphics[width = 1.5 cm, valign=c]{NK-f-21.eps}}
\end{equation}
is written out using standard methods. It is also convenient to split it into two parts: using the nonlocal component
\begin{equation}\label{NK-f-67}
\frac{t_6^2}{24}\frac{L}{16\pi^2}\int_{\mathbb{R}^{3}}\mathrm{d}^3x\,PS^2(x,x)B^2(x)+\frac{\alpha_1(\mathbf{f})\Lambda t_6^2}{48}\int_{\mathbb{R}^{3}}\mathrm{d}^3x\,PS(x,x)B^2(x),
\end{equation}
and also the part that depends only on the background field
\begin{multline}\label{NK-f-68}
\frac{\alpha_2(\mathbf{f})\Lambda^2 t_6^2}{2(5!)}\int_{\mathbb{R}^{3}}\mathrm{d}^3x\,B^2(x)-
\frac{t_6^2}{2(3!5!)}\frac{L}{(16\pi^2)^2}\int_{\mathbb{R}^{3}}\mathrm{d}^3x\,B(x)A_0(x)B(x)-\\-
\frac{\mathrm{I}_2(\Lambda)t_6^2}{48}\int_{\mathbb{R}^{3}}\mathrm{d}^3x\,B^2(x)\bigg(m^2+\frac{t_4B^2(x)}{2}+\frac{t_6B^4(x)}{4!}\bigg).
\end{multline}
Here, for convenience, two new auxiliary numbers have been introduced
\begin{equation}\label{NK-f-69}
\alpha_2(\mathbf{f})=\int_{\mathbb{R}^3}\mathrm{d}^3y\,\Big(R_0^1(y)\Big)^5,
\end{equation}
\begin{equation}\label{NK-f-70}
\mathrm{I}_2(\Lambda)%&\,
=\int_{\mathrm{B}_{1/\sigma}}\mathrm{d}^3x\,\Big(R_0^\Lambda(x)\Big)^4
\Bigg(
\int_{\mathrm{B}_{1/\sigma}}\mathrm{d}^3y\,R_0^\Lambda(x-y)R_0^\Lambda(y)-
\int_{\mathrm{B}_{1/\sigma}}\mathrm{d}^3y\,\Big(R_0^\Lambda(y)\Big)^2
\Bigg)
%\\\noindent&\stackrel{\mathrm{s.p.}}{=}
%\int_{\mathrm{B}_{1/\sigma}}\mathrm{d}^3x\,\Big(R_0^\Lambda(x)\Big)^4
%\Bigg(
%\int_{\mathrm{B}_{1/\sigma}}\mathrm{d}^3y\,R_0(x-y)R_0(y)-
%\int_{\mathrm{B}_{1/\sigma}}\mathrm{d}^3y\,\Big(R_0(y)\Big)^2
%\Bigg)
%\\\noindent&\,=
%\int_{\mathrm{B}_{1/\sigma}}\mathrm{d}^3x\,\Big(R_0^\Lambda(x)\Big)^4\cdot\frac{(-|x|)}{8\pi}
\stackrel{\mathrm{s.p.}}{=}-\frac{L}{2(4\pi)^4},
\end{equation}
where we used the transition $$R_0^\Lambda(x-y)R_0^\Lambda(y)-\Big(R_0^\Lambda(y)\Big)^2 \stackrel{\mathrm{s.p.}}{\to} R_0(x-y)R_0(y)-\Big(R_0(y)\Big)^2 \stackrel{\mathrm{\int}}{\to}\frac{(-|x|)}{8\pi}.$$
Note that \eqref{NK-f-68} can be represented as
\begin{multline}
\eqref{NK-f-68}
\stackrel{\mathrm{s.p.}}{=}
\bigg(\int_{\mathbb{R}^{3}}\mathrm{d}^3x\,B(x)A_0(x)B(x)\bigg)\bigg(-\frac{Lt_6^2}{2^53^25(4\pi)^4}\bigg)+\\+
\mathrm{V}_{0,2}\bigg(\frac{\alpha_2(\mathbf{f})\Lambda^2 t_6^2}{2(5!)}+\frac{Lm^2t_6^2}{2^53(4\pi)^4}\bigg)+
\mathrm{V}_{0,4}\bigg(\frac{Lt_4^{\phantom{1}}t_6^2}{2^63(4\pi)^4}\bigg)+
\mathrm{V}_{0,6}\bigg(\frac{Lt_6^3}{2^83^2(4\pi)^4}\bigg).\label{NK-f-68-1}
\end{multline}

\noindent\textbf{Relation 17.} The last diagram $\mathrm{A}_{24}$ from \eqref{NK-ff-1} contains only the part depending on the background field and can be represented as
\begin{equation}\label{NK-f-71}
	-\frac{\mathbb{H}_0^{\mathrm{sc}}(\Gamma_{40}^3)}{3!(4!)^3}+\kappa_{24}=-\frac{1}{240}{\centering\adjincludegraphics[width = 1.6 cm, valign=c]{NK-f-24.eps}}+\kappa_{24}\stackrel{\mathrm{s.p.}}{=}-\frac{\mathrm{I}_3(\Lambda)}{48}
	\int_{\mathbb{R}^{3}}\mathrm{d}^3x\,\bigg(\Big(t_4+t_6B^2(x)/2\Big)^3-t_4^3\bigg),
\end{equation}
where the constant $\kappa_{24}$ does not depend on the background field and subtracts the excess constant, and the auxiliary integral has the form
\begin{align}\label{NK-f-72}
	\mathrm{I}_3(\Lambda)&\,=\int_{\mathrm{B}_{1/\sigma}}\mathrm{d}^3x
	\int_{\mathrm{B}_{1/\sigma}}\mathrm{d}^3y\,\Big(R_0^\Lambda(x)\Big)^2
	\Big(R_0^\Lambda(x-y)\Big)^2\Big(R_0^\Lambda(y)\Big)^2
\\\nonumber&\stackrel{\mathrm{s.p.}}{=}
\int_{\mathbb{R}^3}\mathrm{d}^3x
\int_{\mathrm{B}_{1/\sigma}}\mathrm{d}^3y\,\Big(R_0^\Lambda(x)\Big)^2
\Big(R_0^\Lambda(x-y)\Big)^2\Big(R_0^\Lambda(y)\Big)^2
\\\nonumber&\stackrel{\mathrm{s.p.}}{=}
L\Bigg(\Lambda\frac{\mathrm{d}}{\mathrm{d}\Lambda}
\int_{\mathbb{R}^3}\mathrm{d}^3x
\int_{\mathrm{B}_{\Lambda/\sigma}}\mathrm{d}^3y\,\Big(R_0^1(x)\Big)^2
\Big(R_0^1(x-y)\Big)^2\Big(R_0^1(y)\Big)^2
\Bigg)\Bigg|_{\Lambda\to+\infty}
\\\nonumber&\,=
\frac{L}{16\pi^2}
\int_{\mathbb{R}^3}\mathrm{d}^3x
\int_{|\hat{y}|=1}\mathrm{d}^2\sigma(\hat{y})\,\Big(R_0(x)\Big)^2
\Big(R_0(x-\hat{y})\Big)^2\equiv\frac{L\alpha_3}{(4\pi)^4}.
\end{align}
Therefore, we obtain
\begin{equation}
\eqref{NK-f-72}\stackrel{\mathrm{s.p.}}{=}
\mathrm{V}_{0,2}\bigg(-\frac{L\alpha_3^{\phantom{1}}t_4^2t_6^{\phantom{1}}}{2^5(4\pi)^4}\bigg)+
\mathrm{V}_{0,4}\bigg(-\frac{L\alpha_3^{\phantom{1}}t_4^{\phantom{1}}t_6^2}{2^6(4\pi)^4}\bigg)+
\mathrm{V}_{0,6}\bigg(-\frac{L\alpha_3^{\phantom{1}}t_6^3}{2^73(4\pi)^4}\bigg).
\label{NK-f-73}
\end{equation}

Let us calculate the value of the coefficient $\alpha_3$:
\begin{equation}
\alpha_3=16\pi^2\int_{\mathbb{R}^3}\mathrm{d}^3x
\int_{|\hat{y}|=1}\mathrm{d}^2\sigma(\hat{y})\,\Big(R_0(x)\Big)^2
\Big(R_0(x-\hat{y})\Big)^2=\frac{1}{2}\int_0^{+\infty} \frac{\mathrm{d}r}{r}\ln\left|\frac{r+1}{r-1}\right|=\int_0^1\frac{\mathrm{d}r}{r}\ln\left(\frac{1+r}{1-r}\right),
\end{equation}
where we have made the transition to spherical coordinates. Note that
\begin{equation}
\ln(1+r)-\ln(1-r)=2\sum_{k=0}^{+\infty}\frac{r^{2k+1}}{2k+1},
\end{equation}
then we get
\begin{equation}
\alpha_3=2\sum_{k=0}^{+\infty}\frac{1}{2k+1}\int_0^1\mathrm{d}r\,r^{2k}=\sum_{k=0}^{+\infty}\frac{2}{(2k+1)^2}=\frac{\pi^2}{4}.
\end{equation}

\section{Conclusion}
\label{NK:sec:dis}

In this paper, singular contributions for the three-dimensional sextic model in the four-loop approximation were studied. It was shown that the result does not depend on nonlocal contributions. Fourth coefficients for the renormalization constants were found.

Note that the renormalization proposed in the paper is implemented within the framework of the minimal subtraction scheme, the so-called MS-scheme. The paper also clearly shows the execution of the $\mathcal{R}$-operation, which is illustrated by the reduction of singular contributions in Relations 1--13.

It can be observed  that the value of the coefficient \eqref{1} for the renormalization constant depends on $\alpha(\mathbf{f})$, $\alpha_1(\mathbf{f})$, and $\alpha_2(\mathbf{f})$, which are given by expressions \eqref{NK-f-z6}, \eqref{NK-f-z7}, and \eqref{NK-f-z8}. Let us consider two special cases.
\begin{itemize}
	\item Let $\mathbf{f}$ has the trivial form $\mathbf{f}=0$, then
	\begin{equation}
	\alpha=\frac{1}{4\pi},\,\,\,\alpha_1=\frac{4}{3(4\pi)^3},\,\,\,\alpha_2=\frac{5}{6(4\pi)^4}.
	\end{equation}
	\item Let $\mathbf{f}(t^2)=1-t$, then the condition of applicability for the regularization from \cite{Iv-2024} is satisfied, and the numbers are 
	\begin{equation}
	\alpha=\frac{1}{2\pi},\,\,\,\alpha_1=\frac{68}{35(4\pi)^3},\,\,\,\alpha_2=\frac{101}{56(4\pi)^4}.
	\end{equation}
	
\end{itemize}

\vspace{2mm}
\noindent\textbf{Acknowledgments.} The author is grateful to A.V. Ivanov for useful discussions and meticulous editorial work.


\begin{thebibliography}{99}
\bibitem{19}
C. G. Bollini, J. J. Giambiagi, \textit{Dimensional Renormalization: The Number of Dimensions as a Regularizing Parameter}, Nuovo Cim. B, \textbf{12}, 20--26 (1972)
\bibitem{555}
G. 't Hooft, M. Veltman, \textit{Regularization and renormalization of gauge fields}, Nucl. Phys. B \textbf{44}, 189--213 (1972)

\bibitem{Bakeyev-Slavnov}
T. Bakeyev, A. Slavnov, \textit{Higher covariant derivative regularization revisited}, Mod. Phys. Lett. A \textbf{11}(19), 1539--1554 (1996)
\bibitem{29-st}
K. V. Stepanyantz, \textit{The Higher Covariant Derivative Regularization as a Tool for Revealing the Structure of Quantum Corrections in Supersymmetric Gauge Theories},  Proc. Steklov Inst. Math. \textbf{309}, 284--298 (2020)
\bibitem{Pauli-Villars}
W. Pauli, F. Villars, \textit{On the Invariant Regularization in Relativistic Quantum Theory}, Rev. Mod. Phys. \textbf{21}(3): 434--444 (1949)

\bibitem{3}
L. D. Faddeev, A. A. Slavnov, \textit{Gauge Fields: An Introduction to Quantum Theory}, Frontiers in Physics \textbf{83}, Addison-Wesley, (1991)
\bibitem{105}
D. I. Kazakov, \textit{Radiative Corrections, Divergences, Regularization, Renormalization, Renormalization Group and All That in Examples in Quantum Field Theory}, JINR, UC-2008-34, 1--91 (2009) arXiv:0901.2208

\bibitem{34}
A. V. Ivanov, N. V. Kharuk, \textit{Quantum equation of motion and two-loop cutoff renormalization for $\phi^3$ model}, Zap. Nauchn. Sem. POMI, \textbf{487}, POMI, St. Petersburg, 2019, 151--166; J. Math. Sci. (N. Y.), \textbf{257}:4 (2021), 526--536, arXiv:2203.04562, doi:10.1007/s10958-021-05500-5
\bibitem{Ivanov-Kharuk-2020}
A. V. Ivanov, N. V. Kharuk, \textit{Two-Loop Cutoff Renormalization of 4-D Yang--Mills Effective Action}, 2020 J. Phys. G: Nucl. Part. Phys. \textbf{48}, 015002, arXiv:2004.05999, 
doi:10.1088/1361-6471/abb939
\bibitem{Ivanov-2022}
A. V. Ivanov, \textit{Explicit Cutoff Regularization in Coordinate Representation}, 2022 J. Phys. A: Math. Theor. \textbf{55}, 495401, arXiv:2209.01783, doi:10.1088/1751-8121/aca8dc
\bibitem{Ivanov-Kharuk-20222}
A. V. Ivanov, N. V. Kharuk, \textit{Formula for two-loop divergent part of 4-D Yang--Mills effective action},  Eur. Phys. J. C \textbf{82}, 997 (2022), arXiv:2203.07131, doi:10.1140/epjc/s10052-022-10921-w
\bibitem{Ivanov-Akac}
P. V. Akacevich, A. V. Ivanov, \textit{On Two-Loop Effective Action of 2D Sigma Model},  Eur. Phys. J. C \textbf{83}, 653 (2023), arXiv:2304.02374, doi:10.1140/epjc/s10052-023-11797-0
\bibitem{Ivanov-Kharuk-2023}
A. V. Ivanov, N. V. Kharuk, \textit{Three-loop divergences in effective action of
4-dimensional Yang--Mills theory with cutoff regularization: $\Gamma_4^2$-contribution}, Zap. Nauchn. Sem. POMI, \textbf{520}, POMI, St. Petersburg, 2023, 162--188, J Math Sci \textbf{284}, 681--699 (2024) doi:10.1007/s10958-024-07379-4
\bibitem{Iv-2024-1}
A. V. Ivanov, \textit{Three-loop renormalization of the quantum action for a four-dimensional scalar model with quartic interaction with the usage of the background field method and a cutoff regularization}, Nucl. Phys. B, \textbf{1006}, 116647 (2024), doi:10.1016/j.nuclphysb.2024.116647, arXiv:2402.14549, https://www.pdmi.ras.ru/preprint/2024/24-02.html
\bibitem{Iv-Kh-2024}
A. V. Ivanov, N. V. Kharuk, \textit{Three-loop renormalization of the quantum action for a five-dimensional scalar cubic model with the usage of the background field method and a cutoff regularization}, Eur. Phys. J. Plus \textbf{139}, 849 (2024) doi:10.1140/epjp/s13360-024-05648-4,
arXiv:2404.07513, https://www.pdmi.ras.ru/preprint/2024/24-05.html
\bibitem{Kh-2024}
N. V. Kharuk, \textit{Three-loop renormalization with a cutoff in a sextic model}, Questions of quantum field theory and statistical physics. Part 30, Zap. Nauchn. Sem. POMI, \textbf{532}, POMI, St. Petersburg, 2024, 273--286 https://mathscinet.ams.org/mathscinet/relay-station?mr=4811943
https://www.mathnet.ru/eng/znsl7462
\bibitem{Iv-2024}
A. V. Ivanov, \textit{An applicability condition of a cutoff regularization in the coordinate representation}, Funktsional. Anal. i Prilozhen., \textbf{59}:1 (2025), 5--17 arXiv:2403.09218, https://www.pdmi.ras.ru/preprint/2024/24-04.html, DOI: https://doi.org/10.4213/faa4221

\bibitem{NK-1}
L. N. Lipatov, \textit{Calculation of the Gell--Mann--Low function in scalar theory with strong nonlinearity}, Sov. Phys. JETP \textbf{44}, 1055--1062 (1976)
\bibitem{NK-2}
R. D. Pisarski, \textit{Fixed points of $(\phi^6)_3$ and $(\phi^4)_4$ theories}, Phys. Rev. D \textbf{28}, 1554--1556 (1983)
\bibitem{NK-3}
W. A. Bardeen, M. Moshe, M. Bander, \textit{Spontaneous Breaking of Scale Invariance and the Ultraviolet Fixed Point in $O(N)$-Symmetric $(\phi^6_3)$ Theory}, Phys. Rev. Lett. \textbf{52}, 1188--1191
(1984)
\bibitem{NK-4}
R. Gudmundsdottir, G. Rydnell, P. Salomonson, \textit{More on $O(N)$-Symmetric $\phi^6_3$ Theory}, Phys. Rev. Lett. \textbf{53}, 2529--2531 (1984)
\bibitem{NK-5}
J. S. Hager, \textit{Six-loop renormalization group functions of $O(n)$-symmetric $\phi^6$-theory and $\epsilon$-expansions of tricritical exponents up to $\epsilon^3$}, J. Phys. A \textbf{35}, 2703-2711 (2002)
\bibitem{NK-6}
J. A. Gracey, \textit{Renormalization of scalar field theories in rational spacetime dimensions}, Eur. Phys. J. C \textbf{80}, 604 (2020)

\bibitem{sk-7}
A. S. Cattaneo, P. Mnev, N. Reshetikhin, \textit{Perturbative Quantum Gauge Theories on Manifolds with Boundary}, Commun. Math. Phys. \textbf{357}, 631--730 (2018) doi:10.1007/s00220-017-3031-6
\bibitem{sk-14}
S. Kandel, \textit{Functorial quantum field theory in the Riemannian setting}, Adv. Theor. Math.
Phys. \textbf{20}(6), 1443--1471 (2016) doi:10.4310/ATMP.2016.v20.n6.a5
\bibitem{sk-16}
S. Kandel, P. Mnev, K. Wernli, \textit{Two-dimensional perturbative scalar QFT and Atiyah-Segal gluing}, Adv. Theor. Math.
Phys. \textbf{25}(7), 1847--1952 (2021) doi:10.4310/ATMP.2021.v25.n7.a5
\bibitem{sksk}
A. V. Ivanov, \textit{Effective actions, cutoff regularization, quasi-locality, and gluing of partition functions}, J. Phys. A: Math. Theor., \textbf{58}, 135401 (2024) doi:10.1088/1751-8121/adc3de, arXiv:2411.13857, https://www.pdmi.ras.ru/preprint/2024/24-11.html

\bibitem{102}
B. S. DeWitt, \textit{Quantum Theory of Gravity. 2. The Manifestly Covariant
	Theory}, Phys. Rev. \textbf{162}, 1195--1239 (1967)
\bibitem{103}
B. S. DeWitt, \textit{Quantum Theory of Gravity. 3. Applications of the Covariant Theory}, Phys. Rev. \textbf{162}, 1239--1256 (1967)
\bibitem{24}
G. ’t Hooft, \textit{The background field method in gauge field theories}, (Karpacz, 1975), Proceedings, Acta Universitatis Wratislaviensis, \textbf{1}, Wroclaw, 345--369 (1976)
\bibitem{25}
L. F. Abbott, \textit{Introduction to the background field method}, Acta Phys. Polon. B, \textbf{13}:1--2, 33--50 (1982)
\bibitem{26}
I. Ya. Aref'eva, A. A. Slavnov, L. D. Faddeev, \textit{Generating functional for the S-matrix in gauge-invariant theories}, TMF, \textbf{21}:3, 311--321 (1974)

\bibitem{nknk-2}
R. Shrock, \textit{Study of the ultraviolet behavior of an $O(N)$ $\phi^6$
theory in $d=3$ dimensions}, Phys. Rev. D \textbf{107}, 096009 (2023)
doi:10.1103/PhysRevD.107.096009
\bibitem{nknk-1}
S. Kvedaraitė, T. Steudtner, M. Uetrecht, \textit{Revisiting the $\phi^6$ Theory in Three Dimensions at Large $N$}, arXiv:2502.07880 (2025)

	
\bibitem{6}
J. C. Collins, \textit{Renormalization: An Introduction to Renormalization, the Renormalization Group and the Operator-Product Expansion}, Cambridge University Press (1984)
\bibitem{7}
O. I. Zavialov, \textit{Renormalized quantum field theory}, Kluwer Academic Publishers, Dodrecht, Boston, (1990)

\bibitem{29}
M. Lüscher, \textit{Dimensional regularisation in the presence of large background fields}, Annals of Physics \textbf{142}, 359--392 (1982)
\bibitem{30-1-1}
A. V. Ivanov, N. V. Kharuk, \textit{Special Functions for Heat Kernel Expansion}, Eur. Phys. J. Plus \textbf{137}, 1060 (2022), arXiv:2106.00294v2, 10.1140/epjp/s13360-022-03176-7

\bibitem{I-R}
A. V. Ivanov, M. A. Russkikh, \textit{Quantum Field Theory on the Example of the Simplest Cubic Model}, J Math Sci \textbf{275}, 306--325 (2023) https://doi.org/10.1007/s10958-023-06683-9

\bibitem{Bog-R}
N. N. Bogolyubov, D. V. Shirkov, \textit{Introduction to the Theory of Quantized Fields}, Willey, New York, 1--620 (1980)
	
\end{thebibliography}
\end{document}